\documentclass[preprint,12pt,authoryear]{elsarticle}
\usepackage{amsmath,amssymb}
\usepackage{graphicx,color}

\journal{XYZ}

\usepackage{lipsum}
\makeatletter
\def\ps@pprintTitle{%
 \let\@oddhead\@empty
 \let\@evenhead\@empty
 \def\@oddfoot{}%
 \let\@evenfoot\@oddfoot}
\makeatother
\begin{document}

\begin{frontmatter}
\title{On solutions of a Boussinesq-type equation with amplitude-dependent nonlinearities: the case of biomembranes}
\author{J\"uri Engelbrecht, Kert Tamm, Tanel Peets}
\address{Centre for Nonlinear Studies, Institute of Cybernetics at Tallinn University of Technology, Akadeemia tee 21, Tallinn 12618, Estonia\\je@ioc.ee, kert@ioc.ee, tanelp@ioc.ee}
\begin{abstract}
Boussinesq-type wave equations involve nonlinearities and dispersion. In this paper a Boussinesq-type equation with amplitude-dependent nonlinearities is presented. Such a model was proposed by Heimburg and Jackson (2005) for describing longitudinal waves in biomembranes and later improved by Engelbrecht et al. (2015) taking into account the microinertia of a biomembrane. The steady solution in the form of a solitary wave is derived and the influence of nonlinear and dispersive terms over a large range of possible sets of coefficients demonstrated. The solutions emerging from arbitrary initial inputs are found using the numerical simulation. The properties of emerging trains of solitary waves waves are analysed. Finally, the interaction of solitary waves which satisfy the governing equation is studied. The interaction process is not fully elastic and after several interactions radiation effects may be significant. This means that for  the present case the solitary waves are not solitons in the strict mathematical sense. However, like in other cases known in solid mechanics, such solutions may be conditionally called solitons. 
\end{abstract}

\begin{keyword}


nonlinearities\sep dispersion\sep solitary waves\sep solitons\sep emergence\sep interactions
\end{keyword}

\end{frontmatter}


\section{Introduction}
The celebrated wave equation which is based on the conservation of momentum, models the motion with a finite speed. In order to account for accompanying physical phenomena, the wave equation must be modified. For conservative systems, the Boussinesq-type equations are widely used. The original Boussinesq equation was derived for surface waves on a fluid layer \citep{Boussinesq1871,Rayleigh1876} but nowadays such equations are used also in solid mechanics \citep{Christov2007}. The main features of Boussinesq-type equations are: (i) bi-directionality (d'Alembert operator); (ii) nonlinearity of any order; (iii) dispersion of any order (presence of space and time derivatives of the fourth order or higher \citep[etc]{Christov2007}. Beside fluid mechanics, there are many studies of such equations derived using various physical assumptions \citep[etc]{Maugin1995,Maugin1999,Porubov2003,Christov2007,Berezovski2013a,Engelbrecht2015}.
In solid mechanics, nonlinearity is caused by the nonlinear stress-strain relation and nonlinear strain tensor, i.e., physical and geometrical nonlinearities are involved (see, for example, \cite{Engelbrecht1997}). The governing equations involve then $\partial u_i/\partial x_j$ type terms ($i,j=1,2,3$), i.e., the displacement gradients enter the model. For example, the simple 1D equation reads
\begin{equation}
	\label{1Dnonlin}
	u_{tt}-c_0^2(1+ku_x)u_{xx}=0,
\end{equation}
where $u=u_1$, $x=x_1$, $c_0$ is the velocity in the unperturbed state and $k$ is the nonlinear parameter. Here and further, the indices $x$ and $t$ denote the differentiation with respect to the indicated variable.  One could say that actually the velocity $c_0$ is calculated like 
\begin{equation}
	\label{NonLinVel}
	c^2=c_0^2(1+ku_x).
\end{equation}

The dispersive effects in solids are due to the geometry \citep{Porubov2003} or due to the microstructure \citep[etc]{Mindlin1964,Berezovski2013a}. Then terms like $u_{xxxx}$, $u_{xxtt}$                            etc. appear in governing equations. The combined action of nonlinear and dispersive effects may give rise to solitary waves \citep[etc]{Christov2007,Maugin2011,Engelbrecht2011}.

During the last decade the interest to mechanical waves in biomembranes has been growing fast \citep[etc]{Heimburg2005,Andersen2009,Appali2012a}. The biomembranes have a special structure, made of lipids \citep{Heimburg2005,Mueller2014} and in this case nonlinear effects are different from that in solids. Based on experimental results, the nonlinearity in biomembranes can be accounted in the velocity like \citep{Heimburg2005}
\begin{equation}
	\label{LipidNonlinVel}
	c^2=c^2_0+pu+qu^2,
\end{equation}
where $p$ and $q$ are coefficients and $u$ is the density change along the axis of the biomembrane. This means that contrary to the gradient-type nonlinearity, the displacement-type nonlinearity appears in governing equations for waves in biomembranes. The Heimburg-Jackson model \citep{Heimburg2005}, improved by \cite{Engelbrecht2015} takes such nonlinearities into account together with dispersive term(s). The governing equation is then of the Boussinesq-type and may lead to the emergence of solitary waves.

In this paper, the improved Heimburg-Jackson model \citep{Engelbrecht2015} is systematically studied in detail needed for describing the possible emergence of solitary waves. After describing the derivation of the governing Boussinesq-type equation (Section 2), the following questions are analysed: (i) deriving the steady solutions to the governing equation (Section 3); (ii) finding the solutions for an arbitrary input (Section 4); (iii) studying the interaction of waves (Section 5). In this way, the existence of solitary waves is shown, the emergence of trains of solitary waves is demonstrated, and finally, the interaction of solitary waves shows whether the solitary waves are solitons in the classical sense. As it is well known, solitons interact with each other elastically without losses like elementary particles and only the phase shifts show the interaction effects \citep{Zabusky1965,Drazin1989,Salupere2002}. In many physical systems the interaction is accompanied by radiation, i.e., the process is not fully elastic. In this case the solitary waves can only conditionally be called solitons. The final remarks are presented in Section 6 where the special features of solutions to this Boussinesq-type equation with displacement-dependent nonlinearities are summarised. The analysis is wider than only the case of biomembranes and includes many combinations of governing parameters. 

\section{Derivation of the governing equation}
The signal propagation in a nerve fibre is a complicated phenomenon. The nerve fibre itself can be modelled as a tube filled with axoplasm and surrounded by the extracellular fluid. The wall of the tube is made of a biomembrane \citep{Debanne2011}. The biomembrane is a very special biological structure made of phospholipids with hydrophobic tails directed to inside of the membrane, i.e., away from the intra- and extracellular fluid \citep{Mueller2014}. In general words, the lipid membrane represents a special biological microstructure with complicated properties. The concentration of ions within and outside of a fibre is different but the ion change can occur through the ion channels. These channels are closed at the rest but can be opened under electrical or mechanical impact \citep{Mueller2014}. 

The electrophysiological model describing the propagation of an electrical signal called the action potential was derived by \cite{Hodgkin1952} and is based on telegraph equations and on opening and closing the ion channels under the electrical impact. However, this model cannot explain all the complex effects in the nerve fibres. Experiments by \cite{Iwasa1980} and \cite{Tasaki1988} have clearly demonstrated the swelling of the surrounding biomembrane and the accompanying heat exchange. This means that an action potential is accompanied also by a mechanical wave in the fibre wall. A mathematical model governing such a wave is proposed by \cite{Heimburg2005,Heimburg2007}. Their model is based on the wave equation, i.e., on the balance of momentum and written in terms of density change $\Delta \rho_A=u$ in the longitudinal direction:
\begin{equation}
	u_{tt}=(c^2u_x)_x.
\end{equation}

Two essential assumptions are made. First, it is assumed that the velocity $c$ of a wave in a circular biomembrane is related to the compressibility of the lipid structure and can be taken as 
\begin{equation}
	c^2=c^2_0+pu+qu^2,
\end{equation}
where $p$ and $q$ are coefficients and $c_0$ is the velocity of the small amplitude sound wave \citep{Heimburg2005}.

The second assumption is to add \emph{ad hoc} higher order term to the governing equation $-hu_{xxxx}$ responsible for dispersion. The governing equation is then
\begin{equation}
	\label{HJ}
	u_{tt}=\left[(c_0^2+pu+qu^2)u_x\right]_x-hu_{xxxx},
\end{equation}
where $h$ is a constant. Equation \eqref{HJ} is a Boussinesq-type equation \citep[see, for example,][]{Christov2007}. \cite{Heimburg2005} have demonstrated that Eq. \eqref{HJ} possesses a solitary pulse-type solution. There are several further studies analysing such solutions \cite[etc]{Heimburg2007,Andersen2009,Appali2012a}. Equation \eqref{HJ} has been improved by \cite{Engelbrecht2015} in order to remove the discrepancy that at higher frequencies the velocities are unbounded. Following the ideas from the solid mechanics \citep{Mindlin1964,Berezovski2013a} and supported by the Lagrangian formalism, the inertial term is added to the governing equation:
\begin{equation}
	\label{EPT}
	u_{tt}=\left[(c_0^2+pu+qu^2)u_x\right]_x-h_1u_{xxxx}+h_2u_{xxtt},
\end{equation}
where $h_1=h$ and $h_2$ are dispersion coefficients.  

\begin{figure}[h]
\includegraphics[width=0.49\textwidth]{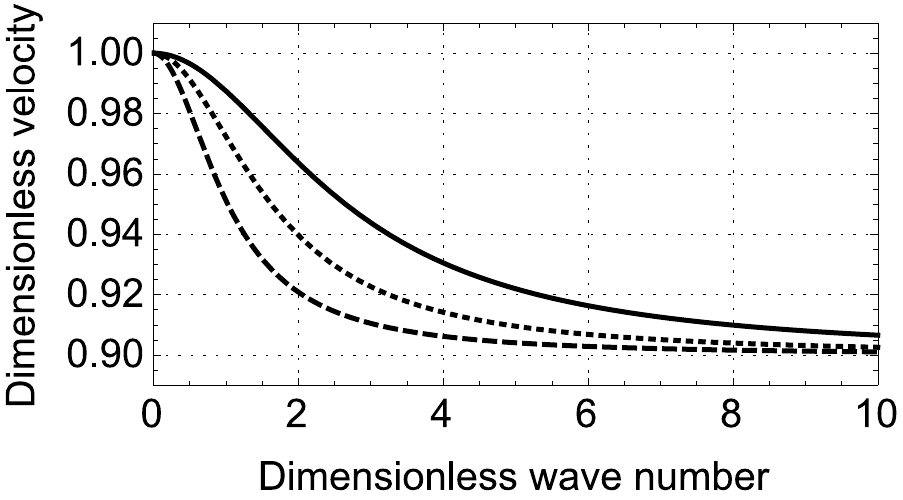}
\includegraphics[width=0.49\textwidth]{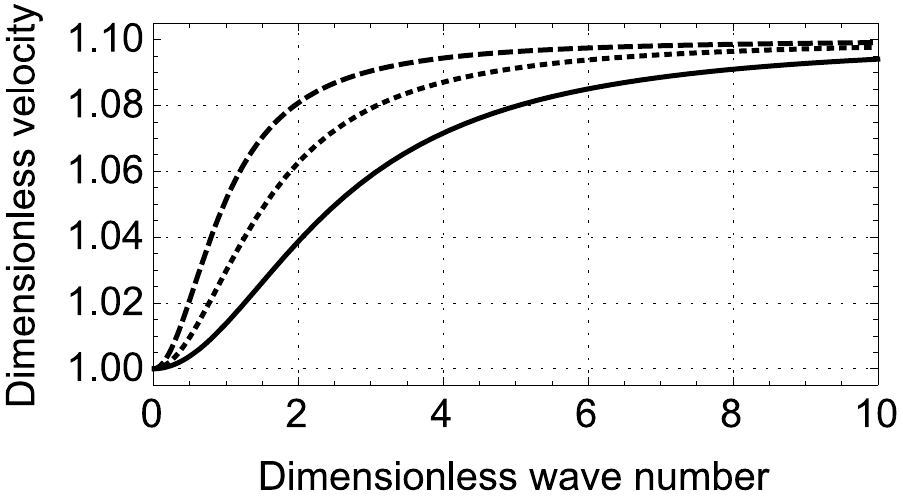}
\caption{Phase speed curves for Eq. \eqref{EPT} in case of $c_1/c_0=0.9$ (left figure) and $c_1/c_0=1.1$ (right figure). $h_2/c_0^2=1$ (dashed), $h_2/c_0^2=0.4$ (dotted line) and $h_2/c_0^2=0.15$ (solid line) in both figures.}
\label{Dispfig}
\end{figure}

The importance of the additional dispersion term $h_2u_{xxtt}$ can be explained with dispersion analysis. It has been shown \citep{Engelbrecht2015} that in case of only one dispersion term $h_1u_{xxxx}$ (Eq.~\eqref{HJ}), the phase velocity is expressed as $c_{ph}^2=c_0^2+h_1k^2$ and it tends to infinity as the wave number $k$ grows. In case of the second fourth order mixed dispersion term $h_2u_{xxtt}$ the propagation velocity is bounded as it can be seen in Fig.~\ref{Dispfig}. The bounding velocity $c_1$ for high frequency harmonics is defined by the ratio of the dispersion coefficients ($c_1^2=h_1/h_2$) and the coefficient $h_2$ is related to the rate of change of the velocity from low frequency to the high frequency domain. Higher valued coefficient $h_2$ means that the transition from the low frequency speeds to the higher frequency speed is more rapid (see Fig.~\ref{Dispfig}). 

From the viewpoint of solid mechanics the importance of the fourth order mixed derivative is not surprising as it is well known that the presence of only spatial derivatives in the governing equation can lead to instabilities \citep{Maugin1999}. Moreover, the mixed fourth order derivative is related to the inertia of the microstructure and it is shown by \cite{Maurin2016} that both dispersive terms arise naturally when all effects involved in wave propagation in solids are considered and this has also been demonstrated experimentally \citep{Maurin2016a}.

The focal point of this paper is the full analysis of Eq. \eqref{EPT}. Further it is convenient to use the dimensionless form of Eq. \eqref{EPT}, which will take the form 
\begin{equation}
	\label{EPTdimensionless}
	U_{TT}=(1+PU+QU^2)U_{XX}+(P+2QU)U^2_x-H_1U_{XXXX}+H_2U_{XXTT},
\end{equation}
where $X=x/l$, $T=c_0t/l$, $U=u/\rho_A$ and $P=p\rho_A/c_0^2$, $Q=q\rho_A^2/c_0^2$. Here $l$ is a certain length, for example, the fibre diameter. 

Equation \eqref{EPTdimensionless} must be solved under initial and boundary conditions formulated in the dependent 
variable $U$.

\section{Steady solutions}

In this section we focus our analysis on undistorted travelling waves in the form
\begin{equation}
	\label{Ansatz}
	V=V(\xi), \quad \xi=X-c T ,
\end{equation}where $V$ is some function and $c$ is dimensionless wave velocity \citep{Ablowitz2011,Drazin1989}. 
Substituting this into Eq. \eqref{EPTdimensionless} we get
\begin{equation}
	\label{EPTode}
	c^2V''=((1+PV+QV^2)V')'-H_1V''''+H_2c^2V''''.
\end{equation}
Integrating Eq.~\eqref{EPTode} twice we get after some rearranging 
\begin{equation}
	\label{EPTodeInt}
	(H_1-H_2c^2)V''=(1-c^2)V+\frac{1}{2}PV^2+\frac{1}{3}QV^3+AV+B,
\end{equation}
where $A$ and $B$ are constants of integration. Since we are looking for solitary wave solutions, then we may add boundary conditions that $V,V',V''\rightarrow 0$ as $X\rightarrow\pm\infty$ and therefore $A,B=0$ \citep{Ablowitz2011,Drazin1989}. Now the Eq.~\eqref{EPTodeInt} is multiplied by $V'$ and integrated to get
\begin{equation}
	\label{PseudoPotPoly}
	(H_1-H_2c^2)(V')^2=(1-c^2)V^2+\frac{1}{3}PV^3+\frac{1}{6}QV^4,	
\end{equation}
which can be rewritten as 
\begin{equation}
	\label{Pseudopot}
	(H_1-H_2c^2)(V')^2=\Phi_{eff}(V),
\end{equation}
where 
\begin{equation}
	\label{Poly}
	\Phi_{eff}(V)=(1-c^2)V^2+\frac{1}{3}PV^3+\frac{1}{6}QV^4
\end{equation}
is a fourth-order `pseudo-potential'. Note that for the classical KdV equation the `pseudo-potential' is of the third order \citep{Drazin1989}.

The existence of solitary waves can be analysed by either investigating the behaviour of the `pseudo-potential' \eqref{Poly} or the phase portrait of Eq. \eqref{PseudoPotPoly}. In case of $H_2=0$ the `pseudo-potential' \eqref{Poly} also applies for the Heimburg-Jackson model \eqref{HJ} and has been analysed by \cite{Lautrup2011} for a particular set of parameters that were determined experimentally and are relevant for the solitary wave propagation in biomembranes ($P<0$, $Q>0$). Here the analysis is more general and the signs of the parameters $P$ and $Q$ are not fixed.

The four zeros of the polynomial \eqref{Poly} are 
\begin{equation}
	\label{zeros}
V_{1,2}=0\quad \text{and}\quad V_{3,4}=\frac{P}{Q}\left(-1\pm\sqrt{1-\frac{(1-c^2)6Q}{P^2}}\right). 
\end{equation}
Double zero at $V_{1,2}=0$ indicates the saddle point, which is minimal requirement for the existence of solitary waves \citep{Ablowitz2011,Drazin1989}. The following analysis can be divided into two parts: the cases of $H_1-H_2c^2>0$, which has also been analysed previously \citep{Peets2016} and $H_1-H_2c^2<0$. Attention is paid to the signs of $P$ and $Q$ which govern the structure of solutions.

(i) $H_1-H_2c^2>0$

\begin{figure}[ht]
\centering
\includegraphics[width=0.45\textwidth]{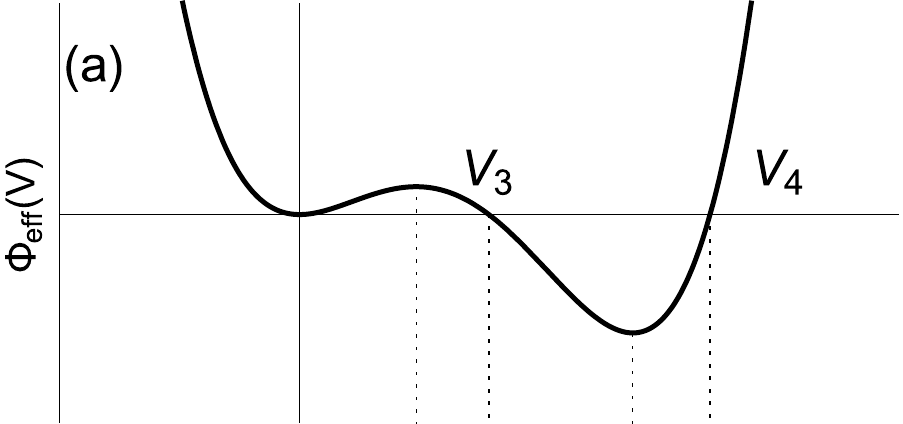}\quad
\includegraphics[width=0.45\textwidth]{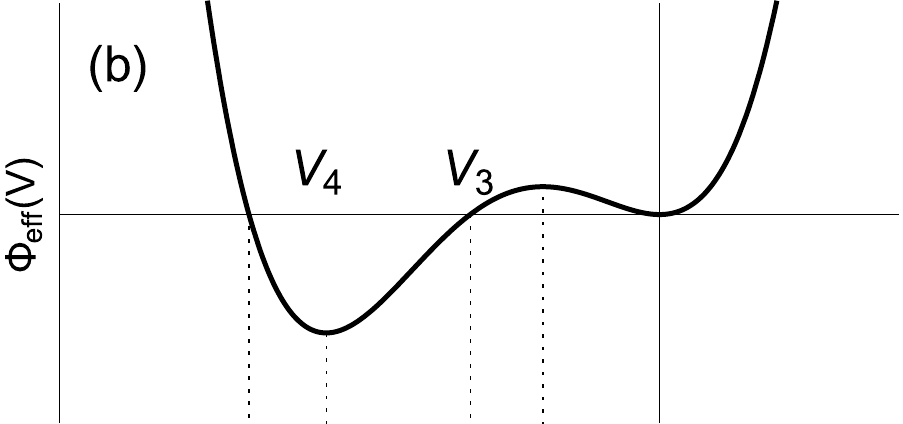}\\ 
\includegraphics[width=0.45\textwidth]{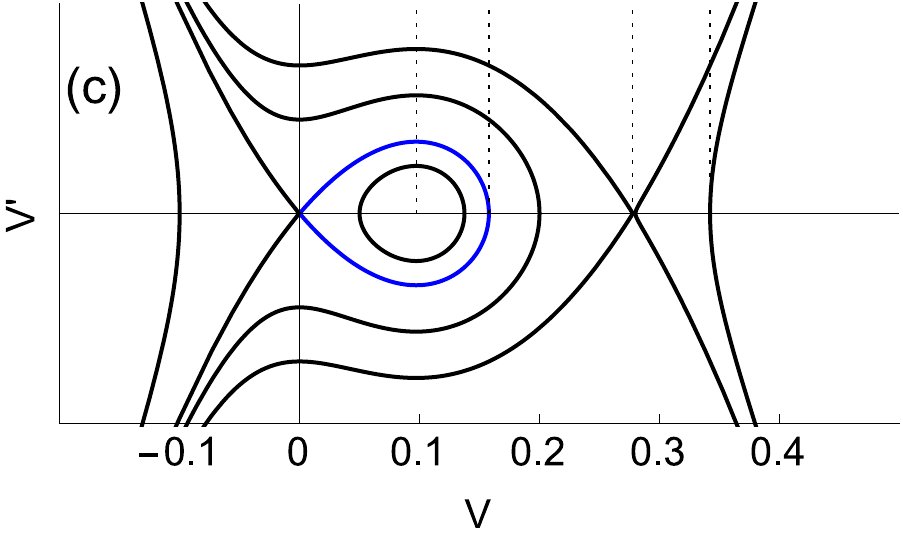}\quad
\includegraphics[width=0.45\textwidth]{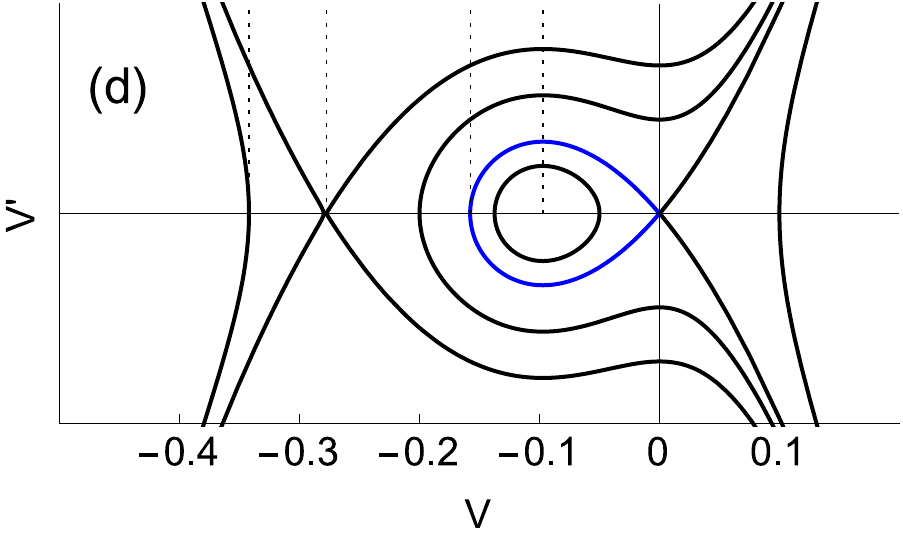}\\
\includegraphics[width=0.45\textwidth]{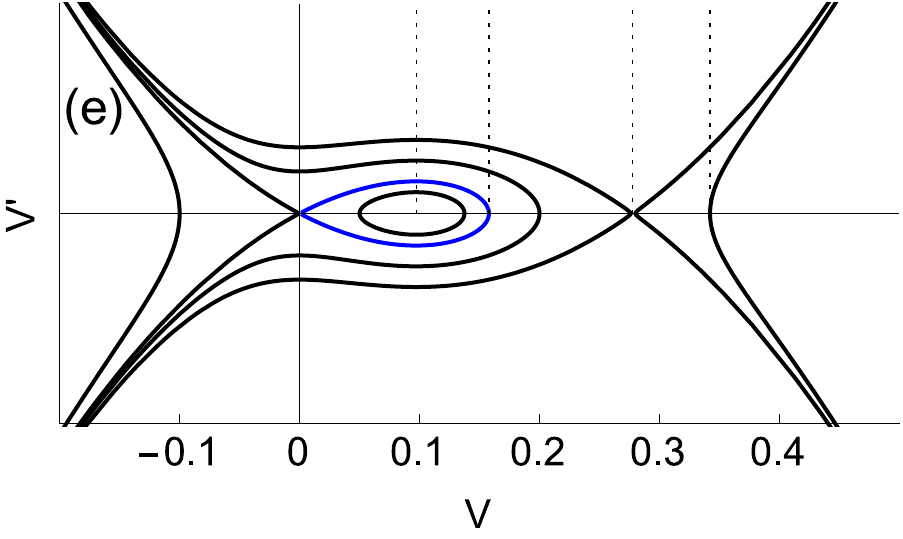}\quad
\includegraphics[width=0.45\textwidth]{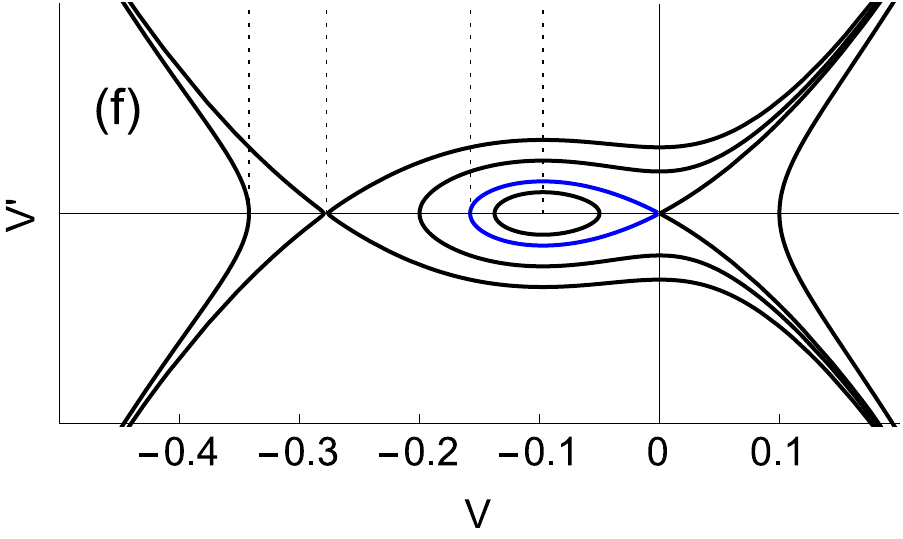}
\caption{Shape of the `pseudopotential' \eqref{Poly} and phase portrait of Eq. \eqref{PseudoPotPoly} in case of $Q>0$. Here $c=0.8$, $Q=40$, $|P|=10$, $H_1=4$ and $H_2=5$. Homoclinic orbit is shown in blue. }
\label{Fig1}
\end{figure}

\emph{The case of $Q>0$}. For this case the analysis is pretty straightforward. It can be deduced from aforementioned restrictions and from Eq.~\eqref{zeros} that in this case the additional condition for the velocity $c$ is
\begin{equation}
	\label{eq15}
	1> |c|>\sqrt{1-\frac{P^2}{6Q}}
\end{equation}
which means that the in case of $Q>0$ and $H_1-H_2c^2>0$ the solitary waves governed by the Eq.~\eqref{EPTdimensionless} will always travel slower than the low frequency sound. This is in good agreement with the actual pulse propagation in biomembranes \citep{Heimburg2005,Lautrup2011}. 

The `pseudopotential' \eqref{Poly} and the phase portrait \eqref{PseudoPotPoly} for this case have been plotted in Fig. \ref{Fig1} for $P<0$ (left column) and for $P>0$ (right column), respectively. The `pseudopotential' \eqref{Poly} has been plotted in the top row and the phase portraits for the case $H_2\neq 0$ in the middle and for the case $H_2=0$ is plotted in the bottom row for reference. 

\begin{figure}[ht]
\centering
\includegraphics[width=0.45\textwidth]{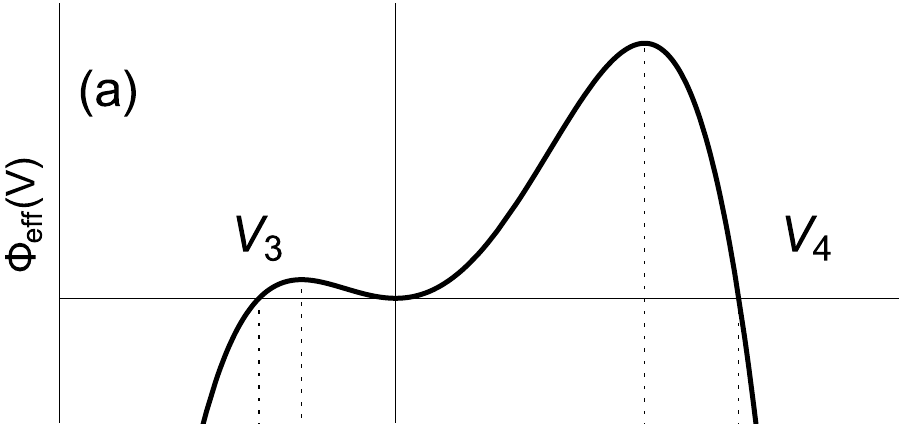}\quad
\includegraphics[width=0.45\textwidth]{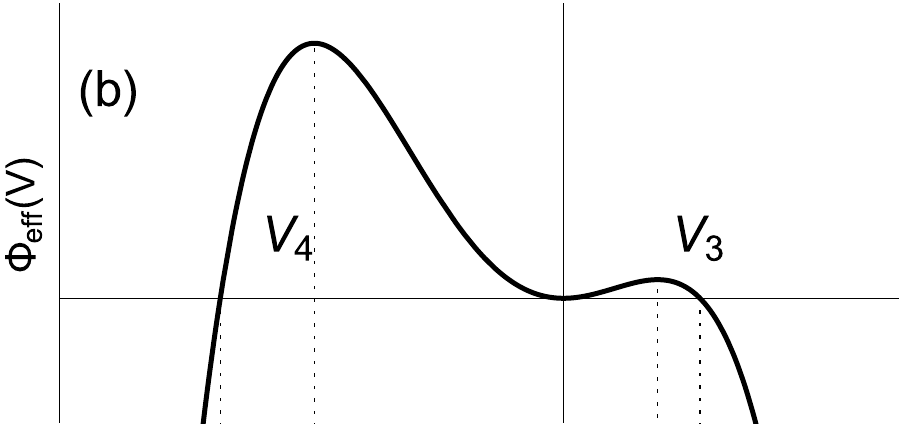}\\
\includegraphics[width=0.45\textwidth]{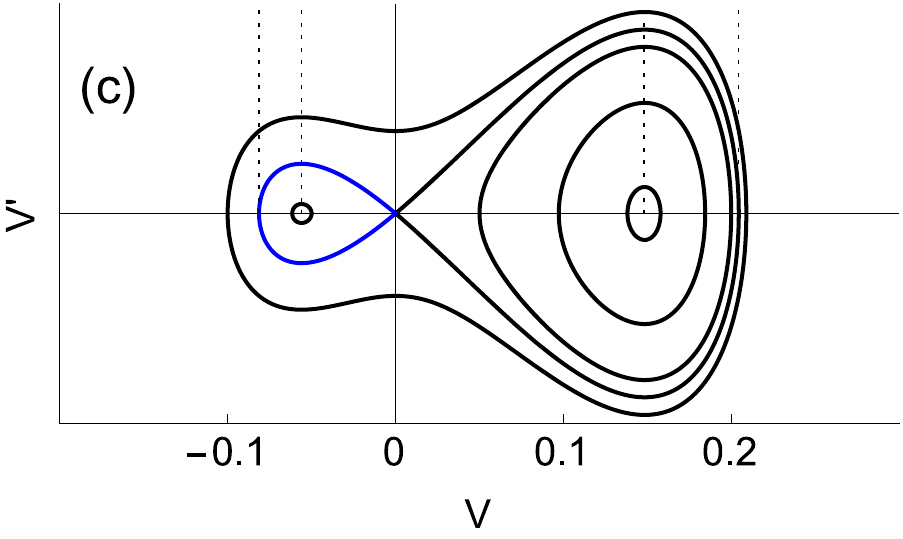}\quad
\includegraphics[width=0.45\textwidth]{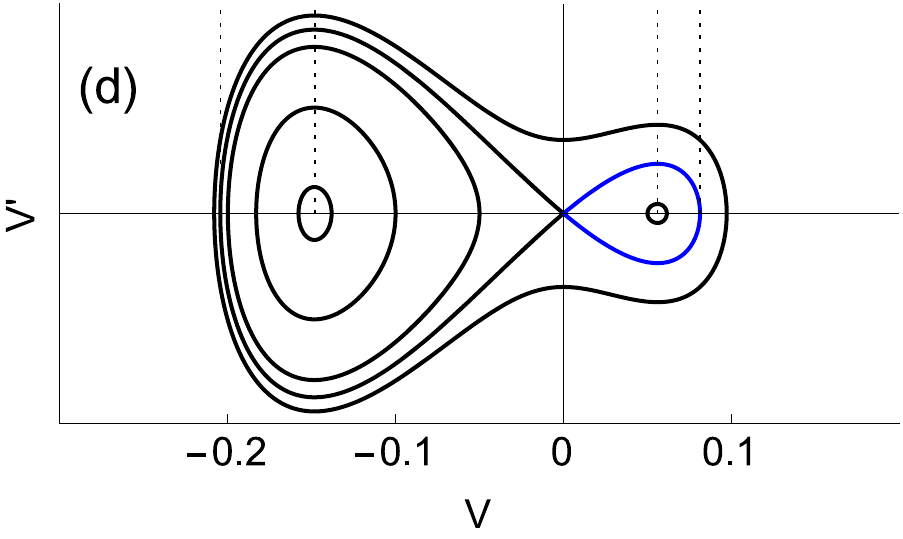}\\
\includegraphics[width=0.45\textwidth]{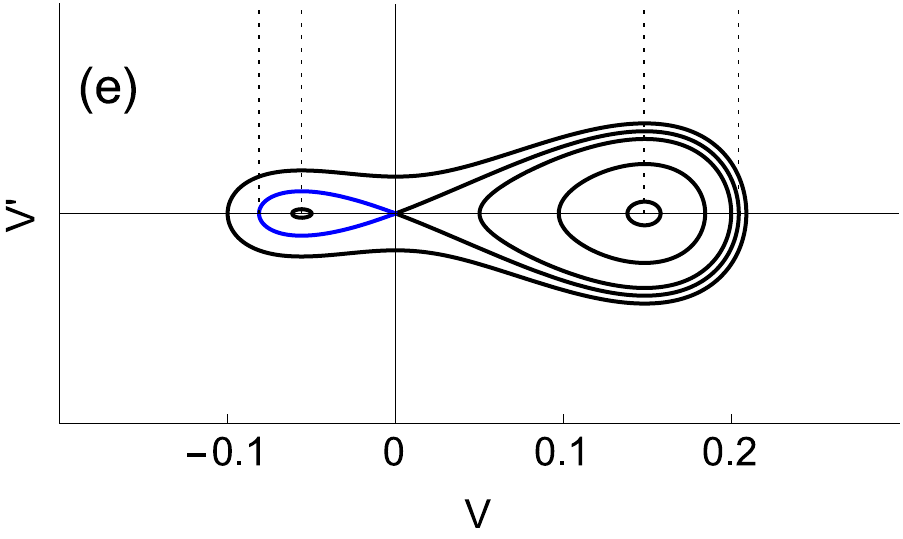}\quad
\includegraphics[width=0.45\textwidth]{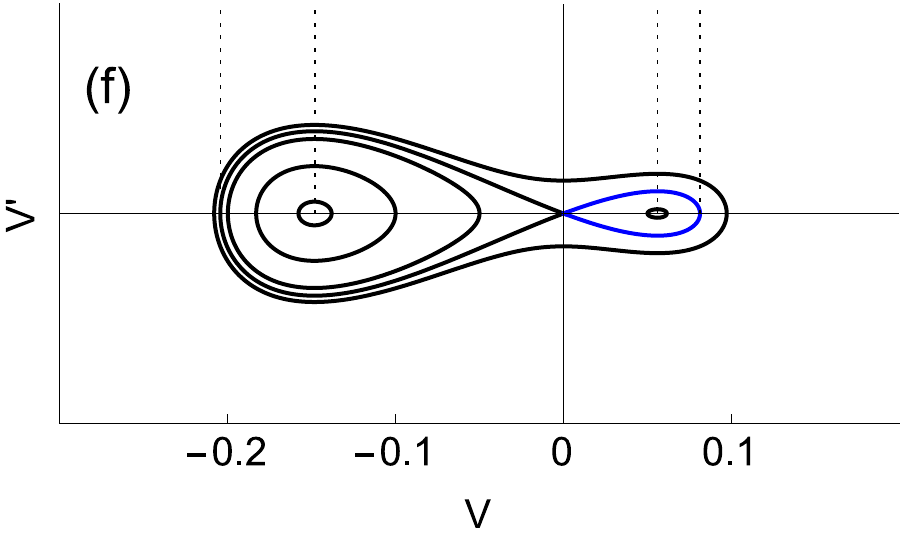}
\caption{Shape of the `pseudopotential' \eqref{Poly} and phase portrait of Eq. \eqref{PseudoPotPoly} in case of $Q<0$. Here $c=0.8$, $Q=-130$, $|P|=8$, $H_1=4$ and $H_2=5$. Homoclinic orbit is shown in blue.}
\label{Fig2}
\end{figure}

The existence of solitary wave solution requires that $\Phi_{eff}(V)$ has a local minimum at $V=0$ with at least one local maximum next to it (Figs \ref{Fig1}a,b). Alternatively one can study the phase portrait (Figs~\ref{Fig1}c,d): solitary wave solutions exist when a saddle point and a homoclinic orbit exists (shown in blue). The amplitude of the solitary wave in both cases is determined by $V_3$. It is clear that while the magnitude of the amplitude of a solitary wave depends on the ratio of the parameters $P$ and $Q$ together with the velocity $c$, the sign of the amplitude is determined only by the parameter $P$: in case of $P<0$ positive solitary wave emerges and in case of $P>0$ the amplitude will be negative. It can also be shown that higher values of $c$ result in lower amplitudes meaning that the lower amplitude solitary waves travel faster as it has been shown earlier \citep{Peets2015,Tamm2015}.

 \emph{The case of $Q<0$} is shown in Fig.~\ref{Fig2}, where it can be seen that the behaviour of the `pseudopotential' and the phase portrait is significantly different from the case of $Q>0$, $H_1-H_2c^2>0$. Although solitary wave solutions are possible in the region where $\Phi_{eff}(V)>0$, only amplitude $V_3$ is realised, because the local maximum between $V_3$ and $V=0$ is closer to the saddle point \citep[cf.][]{Dauxois2006}. Also, in this case the speed of the solitary wave can have any value between zero and $1$ meaning that also in this case the solitary wave travels slower than the speed of the low frequency sound. As in case of $Q>0$, here also the magnitude and the sign of the amplitude is determined by the parameters $P$, $Q$ and $c$ and the higher velocities $c$ result in lower amplitudes.

\begin{figure}[ht]
\centering
\includegraphics[width=0.45\textwidth]{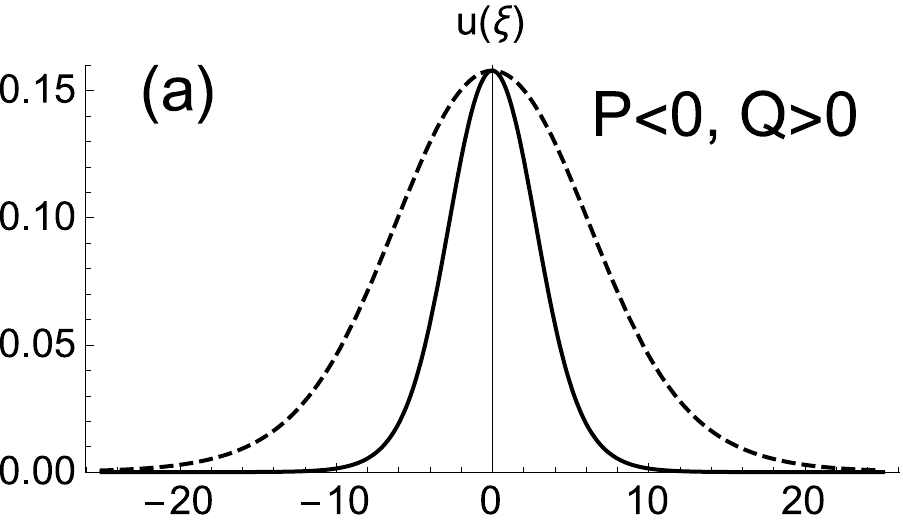}\quad\quad\quad
\includegraphics[width=0.45\textwidth]{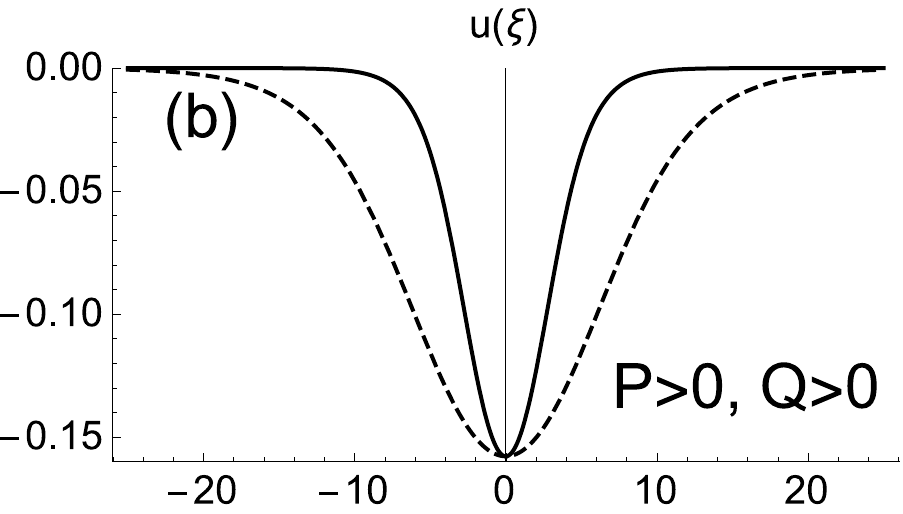}\\
\includegraphics[width=0.45\textwidth]{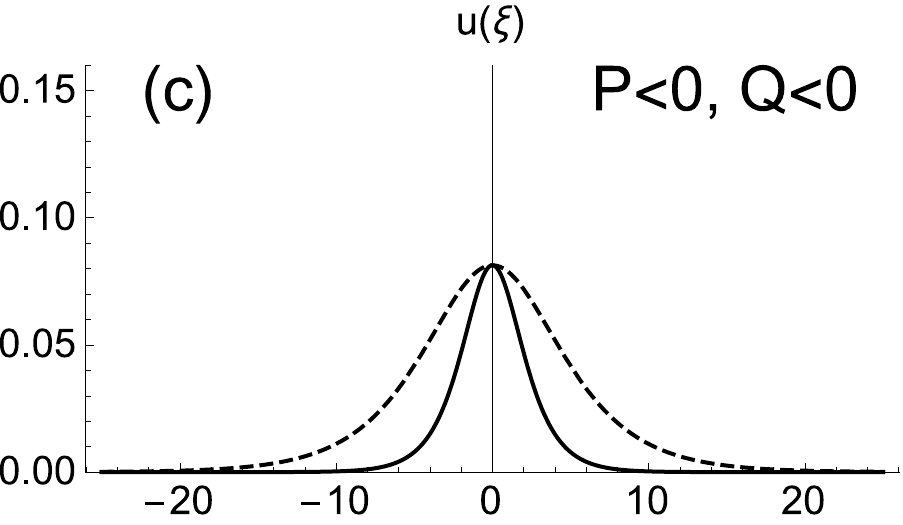}\quad\quad\quad
\includegraphics[width=0.45\textwidth]{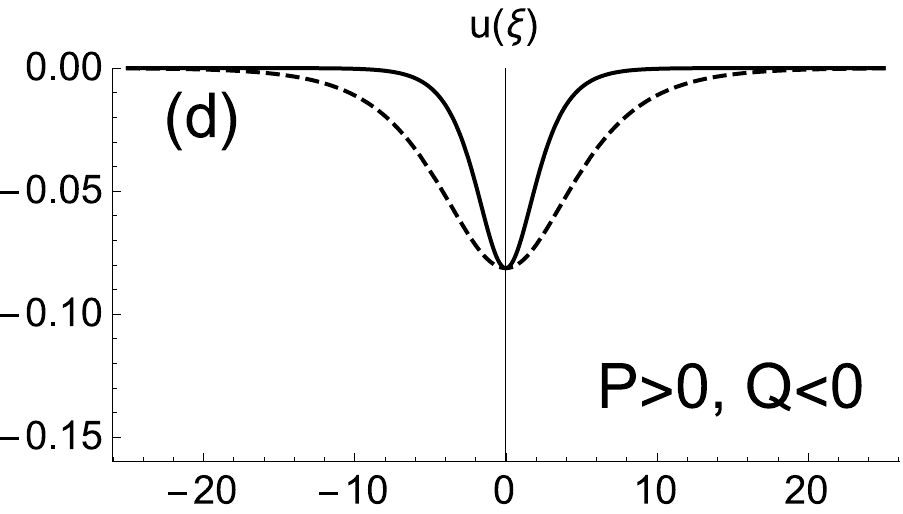}
\caption{Solitary wave solutions of Eq. \eqref{EPTdimensionless} in case of $H_2\neq 0$ (solid line) and $H_2=0$ (dashed line). Here $|P|=16$,$|Q|=80$ (top row) and $|P|=8$,$|Q|=130$ (bottom row);  $c=0.8$, $H_1=2$, $H_2=5$ for all plots.  }
\label{Fig3}
\end{figure}
\begin{figure}[ht]
\centering
\includegraphics[width=0.45\textwidth]{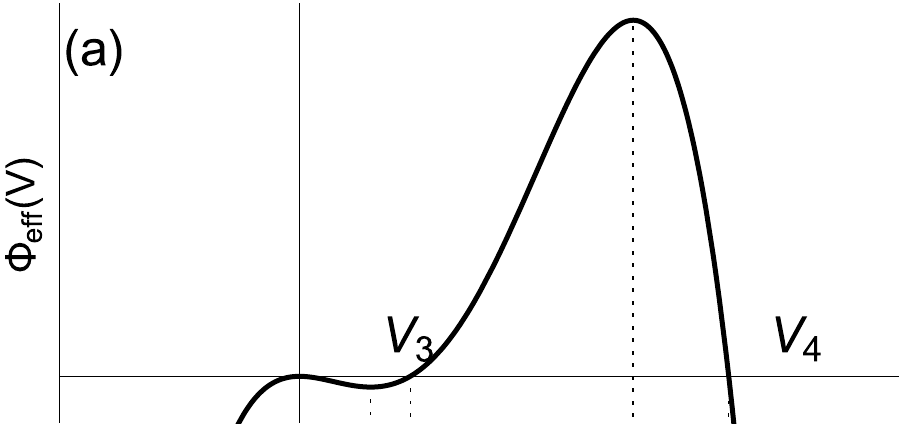}\quad
\includegraphics[width=0.45\textwidth]{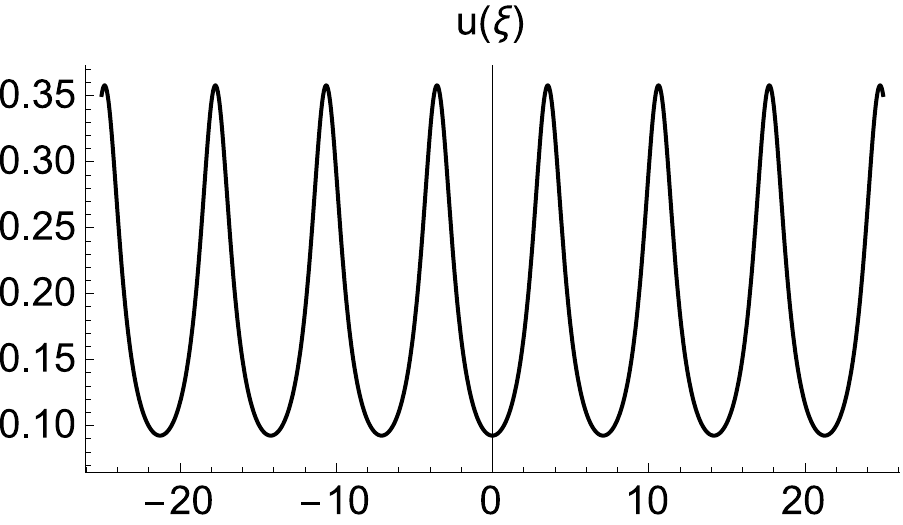}\\
\includegraphics[width=0.45\textwidth]{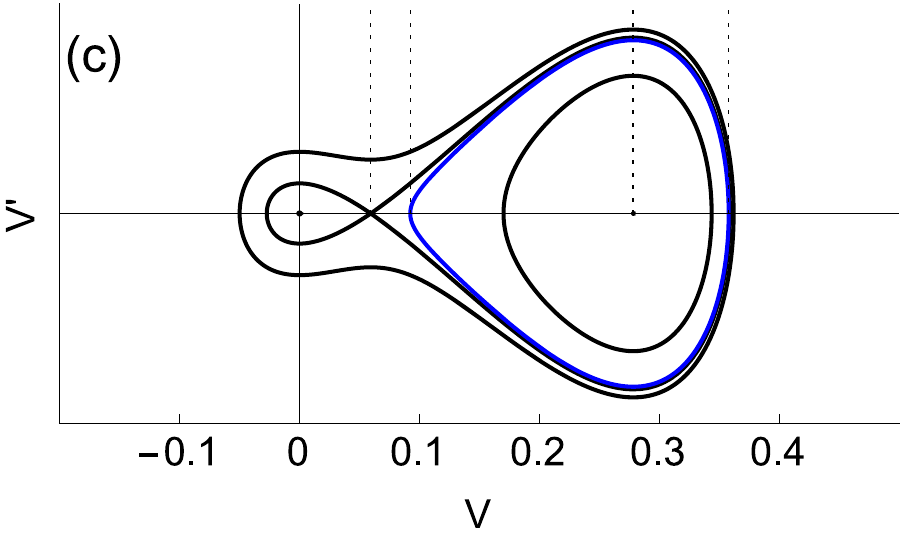}\quad
\includegraphics[width=0.45\textwidth]{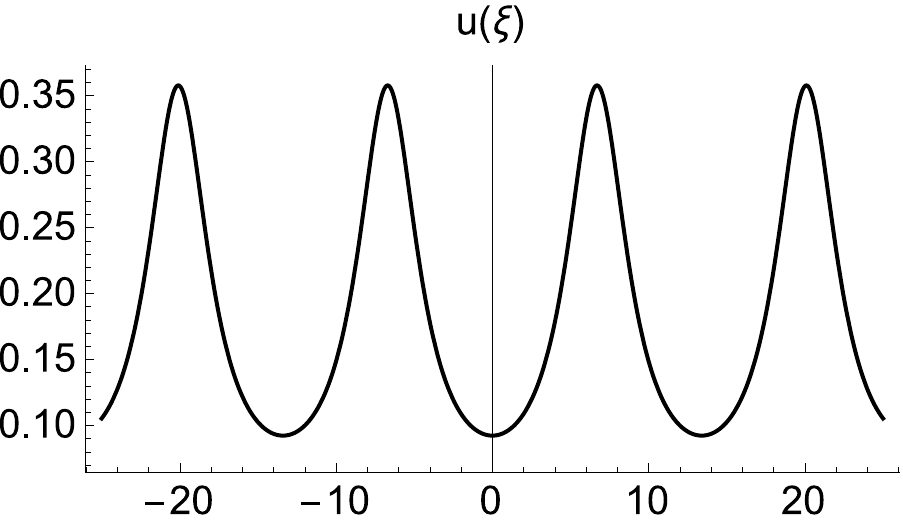}
\caption{Emergence of a periodic wave in case of $Q<0$, $H_1-H_2c^2>0$ and $c>1$: the shape of the `pseudopotential' (a) and the corresponding phase portrait (c). For (b) the parameters are $c=1.2$, $Q=-80$, $P=18$, $H_1=2$, $H_2=1$. For (d) $H_2=20$; other parameters are same as in (b). Stable orbit is shown in blue.}
\label{joon1}
\end{figure}
\begin{figure}[h]
\includegraphics[width=0.45\textwidth]{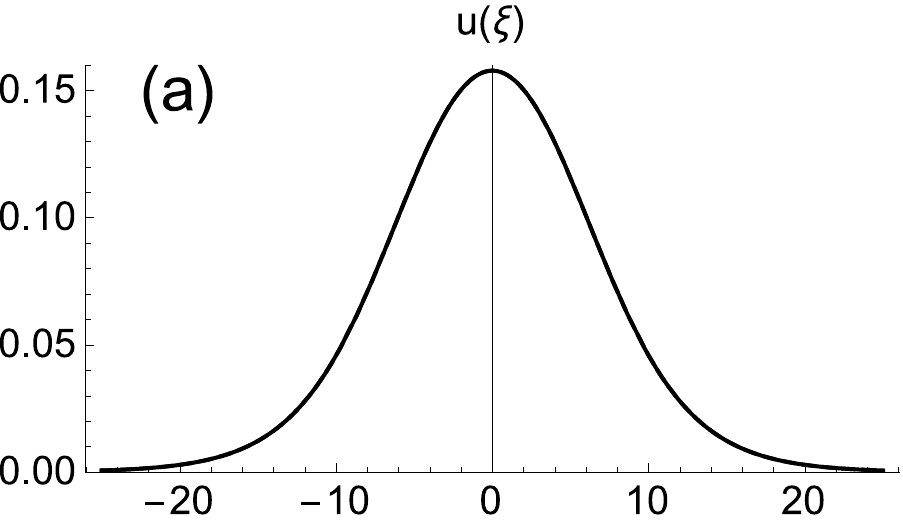}
\includegraphics[width=0.45\textwidth]{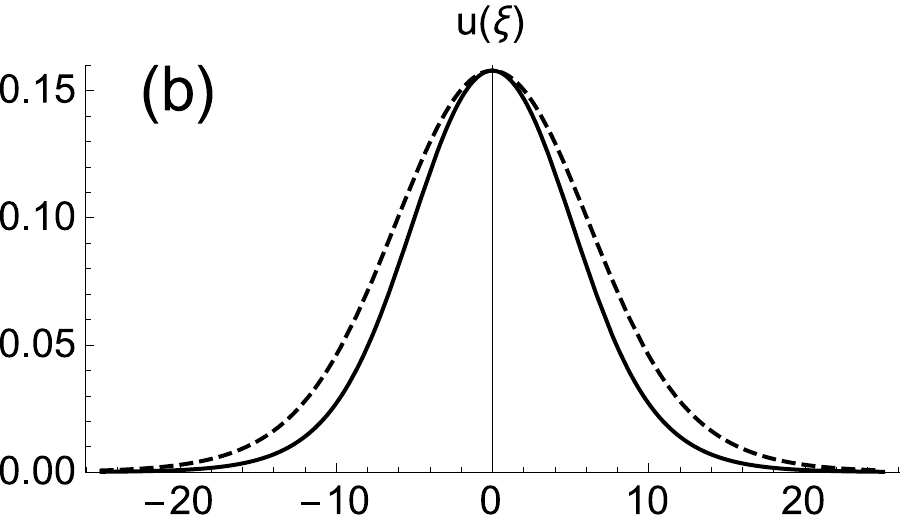}\\
\includegraphics[width=0.45\textwidth]{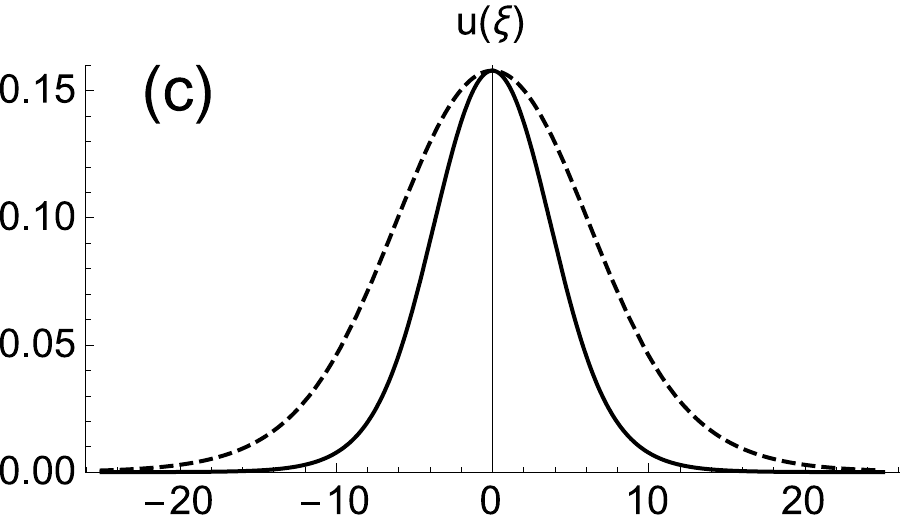}
\includegraphics[width=0.45\textwidth]{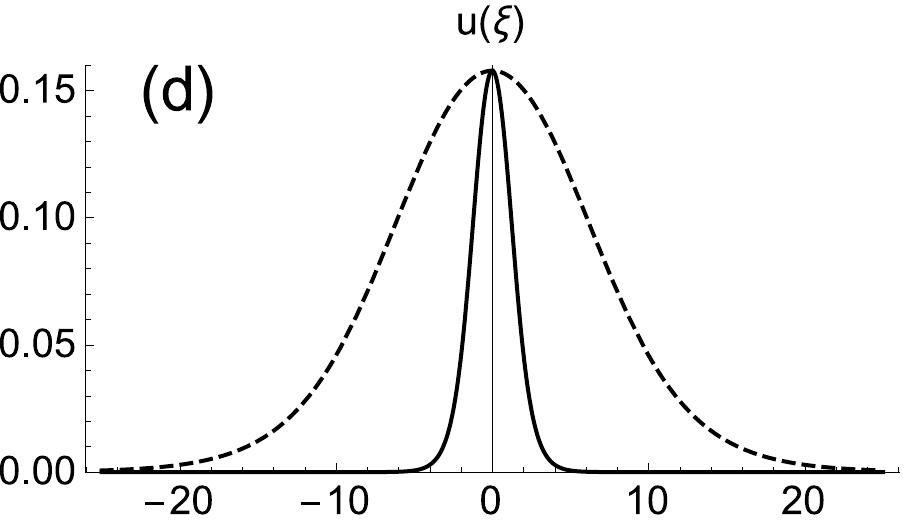}
\caption{The effect of the second dispersive term $H_2U_{XXTT}$ on the width of a solitary wave. Here $P=-10$, $Q=40$, $H_1=2$; (a)$H_2=0$, (b)$H_2=2$, (c)$H_2=4$ and (d)$H_2=6$.}
\label{solitonwidth}
\end{figure}

Recalling that the analytical solution of Eq.~\eqref{EPTdimensionless} is \citep{Peets2015a,Peets2016}
\begin{equation}
	\label{HJsoliton}
	u(\xi)=\frac{6(c^2-1)}{P(1+\sqrt{1+6Q(c^2-1)/P^2}\cosh (\xi\sqrt{(1-c^2)/(H_1-H_2c^2)}) },
\end{equation}
where $\xi=X-c T$ and $c$ is the velocity of the solitary wave, the solitary wave solutions for the given cases are plotted in Fig. \ref{Fig3} where the solid lines represent the case of $H_2\neq 0$ and the dashed represents the case of $H_2=0$, which are the original Heimburg-Jackson equation \eqref{HJ}.

Although solitary waves in case of $Q<0$ and $H_1-H_2c^2>0$ only exist when $0<c<1$, periodic solutions to Eq.~\eqref{PseudoPotPoly} also exist when
\begin{equation}
	1<|c|<\sqrt{1+\left|\frac{P^2}{6Q}\right|}.
\end{equation}
This case is demonstrated in Fig.~\ref{joon1} where it can be seen that in this case `pseudopotential' \eqref{Poly} is positive between the points $V_3$ and $V_4$ and a stable orbit exists (shown in blue) which means an existence of a periodic solution. What is interesting is that the phaseportrait in this case looks similar to Fig.~\ref{Fig2}c only it has been slightly shifted to the right and the higher amplitude part is realised.

It can also be seen in Figs~\ref{Fig1}, \ref{Fig2} and \ref{Fig3} that in the case of the second dispersion coefficient $H_2$ more localised solution is obtained: the greater value of the quantity $V'$ means the steeper slope (and hence the smaller width) of the solitary wave. The effect of the dispersive term $H_2$ on the width of a solitary wave is demonstrated in Fig.~\ref{solitonwidth}, where it can be seen that higher values of $H_2$ result in more localised solutions.

(ii) $H_1-H_2c^2<0$

\begin{figure}[ht]
\centering
\includegraphics[width=0.45\textwidth]{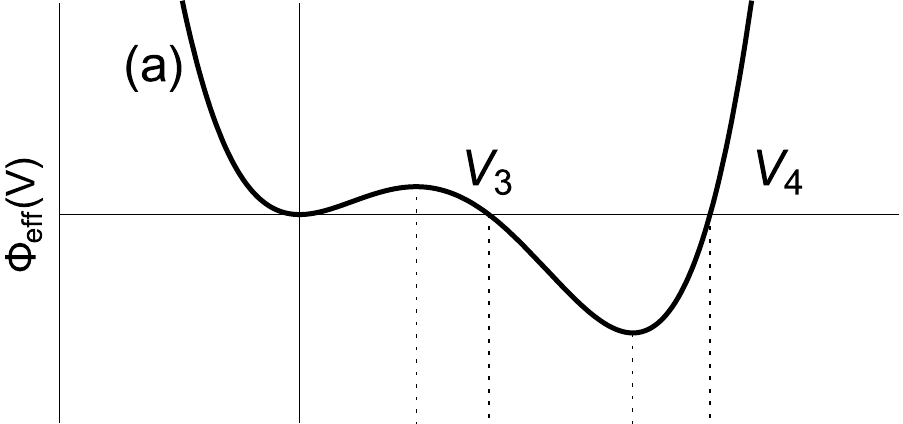}\quad
\includegraphics[width=0.45\textwidth]{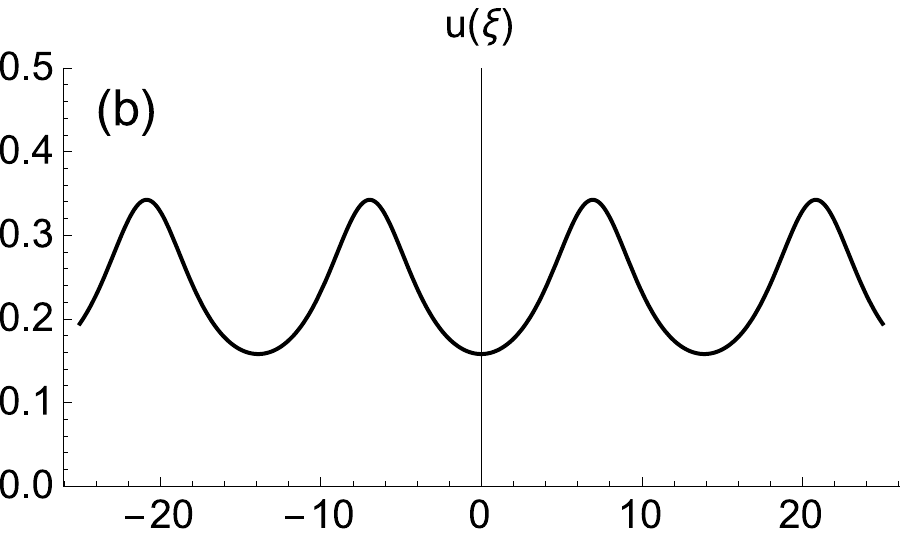}\\
\includegraphics[width=0.45\textwidth]{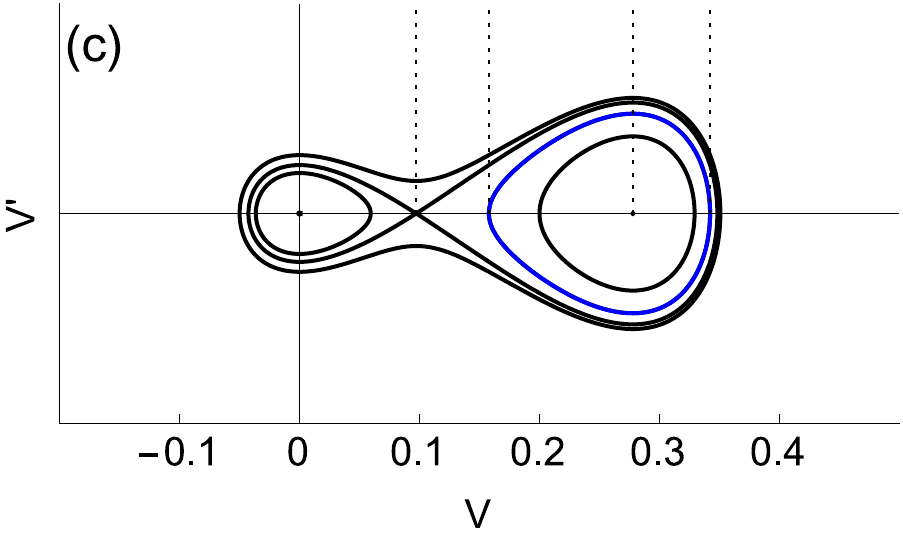}\quad
\includegraphics[width=0.45\textwidth]{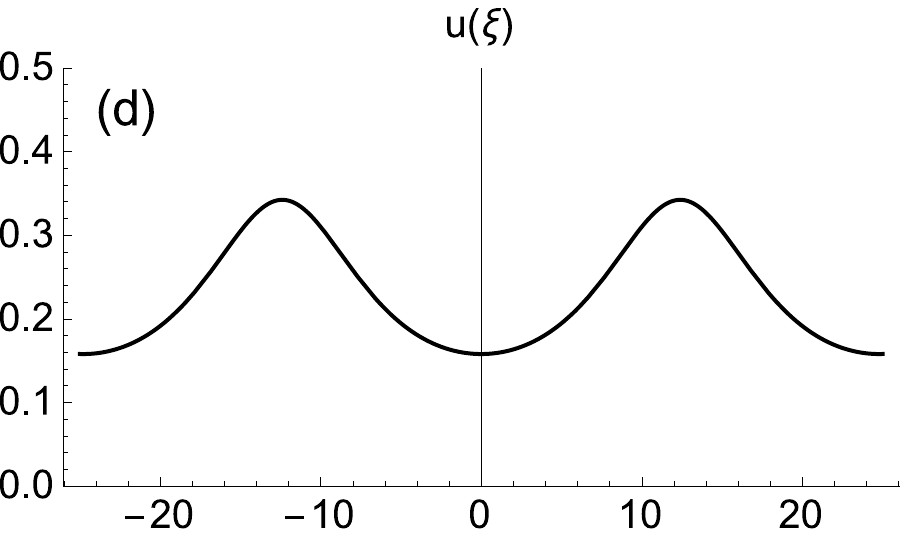}
\caption{Emergence of a periodic wave in case of $H_1-H_2c^2<0$: the shape of the `pseudopotential' (a) and the corresponding phase portrait (c). For (b) the parameters are $c=0.8$, $Q=40$, $P=-10$, $H_1=4$, $H_2=9$. For (d) $H_2=20$; other parameters are same as in (b). Stable orbit is shown in blue.}
\label{Fig4}
\end{figure}

Since Eq.~\eqref{PseudoPotPoly} can be thought of as conservation of `pseudoenergy', then if $H_1-H_2c^2<0$ then also $\Phi_{eff}$ has to be negative. In case of $Q>0$ the condition $H_1-H_2c^2<0$ means that  periodic solutions emerge even in the case of $c<1$ as it is seen in Fig.~\ref{Fig4}. Here also the periodic solution oscillates between the values $V_3$ and $V_4$ (shown in blue in Fig.~\ref{Fig4}), which corresponds to the region where $\Phi_{eff}(V)<0$ as it can be seen in Fig.~\ref{Fig1}a. Similar result is obtained when $P>0$, only with negative amplitude. 
\begin{figure}[ht]
\centering
\includegraphics[width=0.45\textwidth]{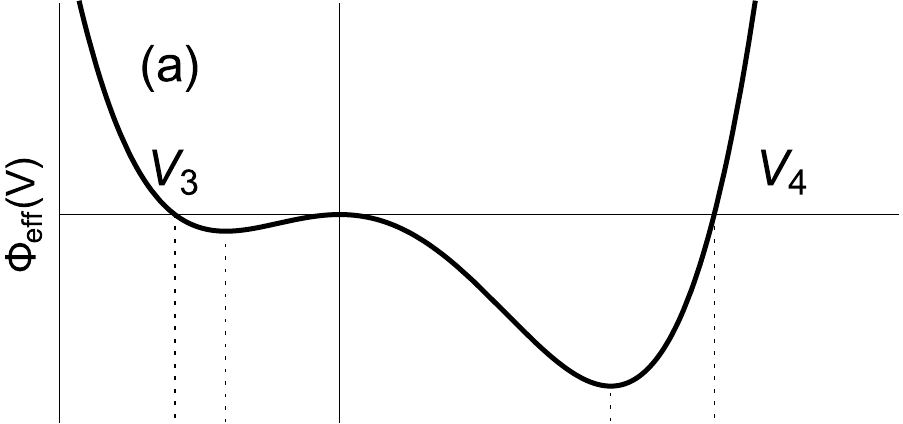}\quad
\includegraphics[width=0.45\textwidth]{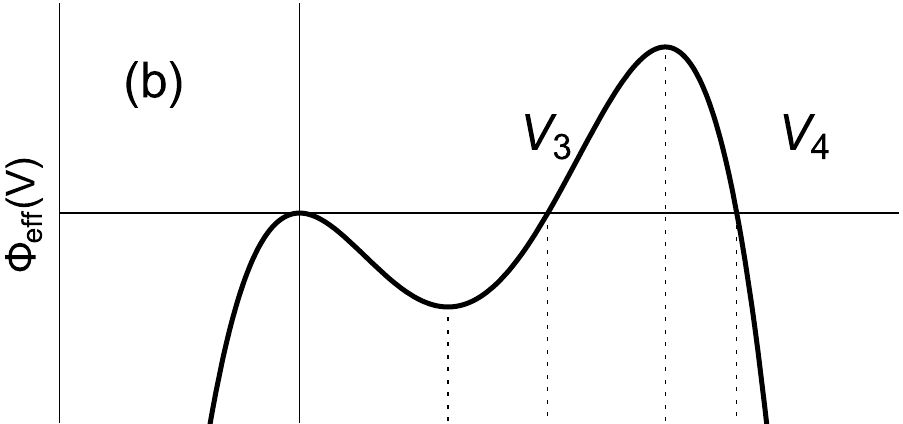}\\
\includegraphics[width=0.45\textwidth]{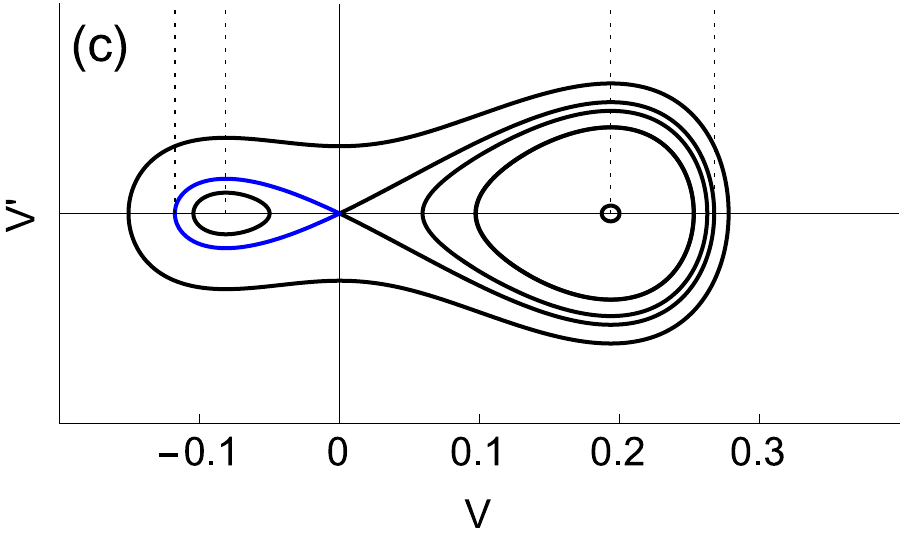}\quad
\includegraphics[width=0.45\textwidth]{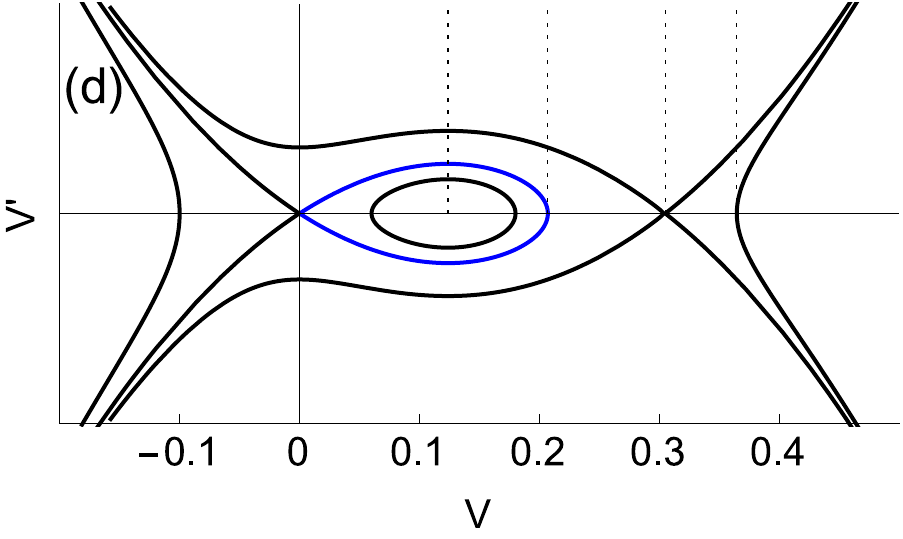}
\caption{Existence of solitary waves in case of $H_1-H_2c^2<0$ and $c>1$. Here $c=1.1$, $Q=40, P=-3$, $H_1=4$ and $H_2=5$ (left column) and $c=1.2$, $Q=-35, P=10$, $H_1=4$ and $H_2=5$ (left column). Homoclinic orbit is shown in blue. }
\label{joon2}
\end{figure}

In case of $c>1$ and $Q>0$ the `pseudopotential' will only have regions $\Phi_{eff}<0$ and in case of $H_1-H_2c^2<0$ a solitary wave with amplitude $V_3$ exist (Fig.~\ref{joon2}a,c). In case of $c>1$ and $Q<0$, a solitary wave with amplitude $V_3$ exists (Fig.~\ref{joon2}b,d). Like in previous cases the structure of the phase portrait depends on the sign of the coefficient $Q$ and the sign of $V_3$ (amplitude of the solution) depends on the sign of the coefficient $P$. Unlike in case of $H_1-H_2c^2>0$ where smaller amplitude solitary waves travel faster, in case of $H_1-H_2c^2<0$ the higher amplitude waves travel faster.  Also recall that if $H_1-H_2c^2>0$ then periodic solution exists with the same coefficients(see Fig.~\ref{joon1}).

Last case we mention is $H_1-H_2c^2<0$, $c<1$ and $Q<0$ when no solutions exist.

\section{Solutions emerging from arbitrary initial conditions}

In the present paper we use the pseudospectral method (PSM) to solve the governing equation \eqref{EPTdimensionless} under localised initial conditions demonstrating the influence of parameters $P, Q, H_1$ and $H_2$ on the evolution of solutions. The PSM is a well established method which is used for solving PDE's and ODE's on regular basis. The advantages and disadvantages of the PSM are well explored in the literature \citep{ForenbergSloan94,Fornberg1998}. Here two points are worth of highlighting: (i) the PSM requires one to use periodic boundary conditions, (ii) the governing equations have to be in a suitable form for applying the PSM with time derivatives on the left hand side and spatial derivatives on the right hand side of the equation. The first point is not a problem, however, taking a look at Eq. \eqref{EPTdimensionless} it is evident that we have a mixed partial derivative term $U_{XXTT}$. We use a change of variables for transforming the governing equation \eqref{EPTdimensionless} for allowing the application of the PSM \citep{Engelbrecht2015,Peets2015,Tamm2015}. The basic idea of the PSM is to find the spatial derivatives by making use of the properties of the Fourier transform and then solve the resulting ODE with respect to time derivate by making use of the commonly available schemes for numerical solving of the ODE's. 

For initial and boundary conditions for the systematic analysis we use a pulse--type localised initial condition in the form
of  $\mathrm{sech}^{2}$-type profile:
$
U(X,0) = U_{o} \mathrm{sech}^2 B_{o} X,
\;
U(X,T) = U (X + 2 k m \pi,T),
\;
m = 1,2,\ldots ,
$
where $k=12$, i.e., the total  length of the spatial period is $24\pi$.
Here the amplitude and the width of the initial pulse are $U_{o} = 1$ and {$B_{o} = 1$}. The initial phase velocity is $U(X,0)_T=0$ meaning that the initial condition splits into two pulses propagating in the opposite directions. Some examples are provided using differentt combination of parameters in which case the used parameters are noted separately. 

The calculations are carried out with the Python package SciPy \citep{SciPy} with Python interface to the ODEPACK FORTRAN code \citep{ODE} for the ODE solver.

\begin{figure}[ht]
\includegraphics[width=0.95\textwidth]{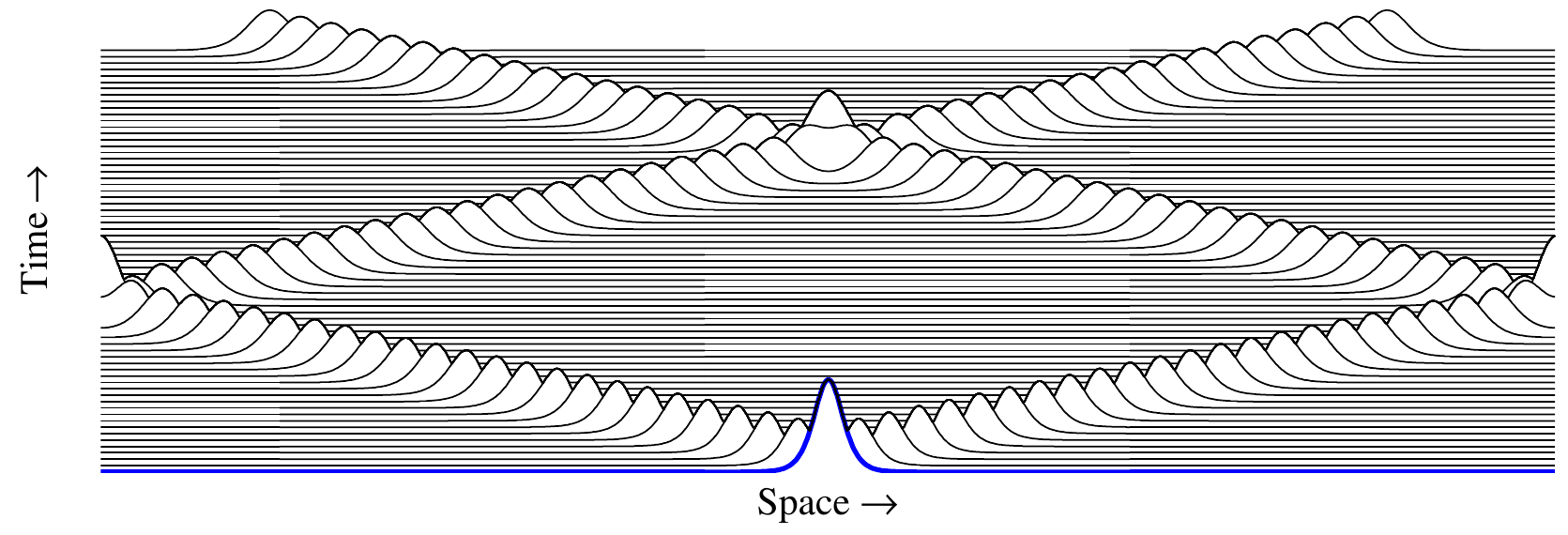}
\caption{Propagation of a typical solution of Eq.~\eqref{EPTdimensionless}. Balanced dispersion case. }
\label{typsol}
\end{figure}

The typical solution of Eq.~\eqref{EPTdimensionless} can be seen in Fig.~\ref{typsol}. Under the arbitrary initial condition the initial disturbance splits into two equal wave structures propagating in opposing directions (initial velocity was zero). In the balanced dispersion case the emerging waveprofiles are of the solitary wave type and very stable even through multiple interactions. 
%Other possibilities are oscillatory structures and trains of solitary waves or some combination of these depending on model parameters and the initial condition characteristics. 

\begin{figure}[ht]
%\begin{centering}
\includegraphics[width=0.49\textwidth]{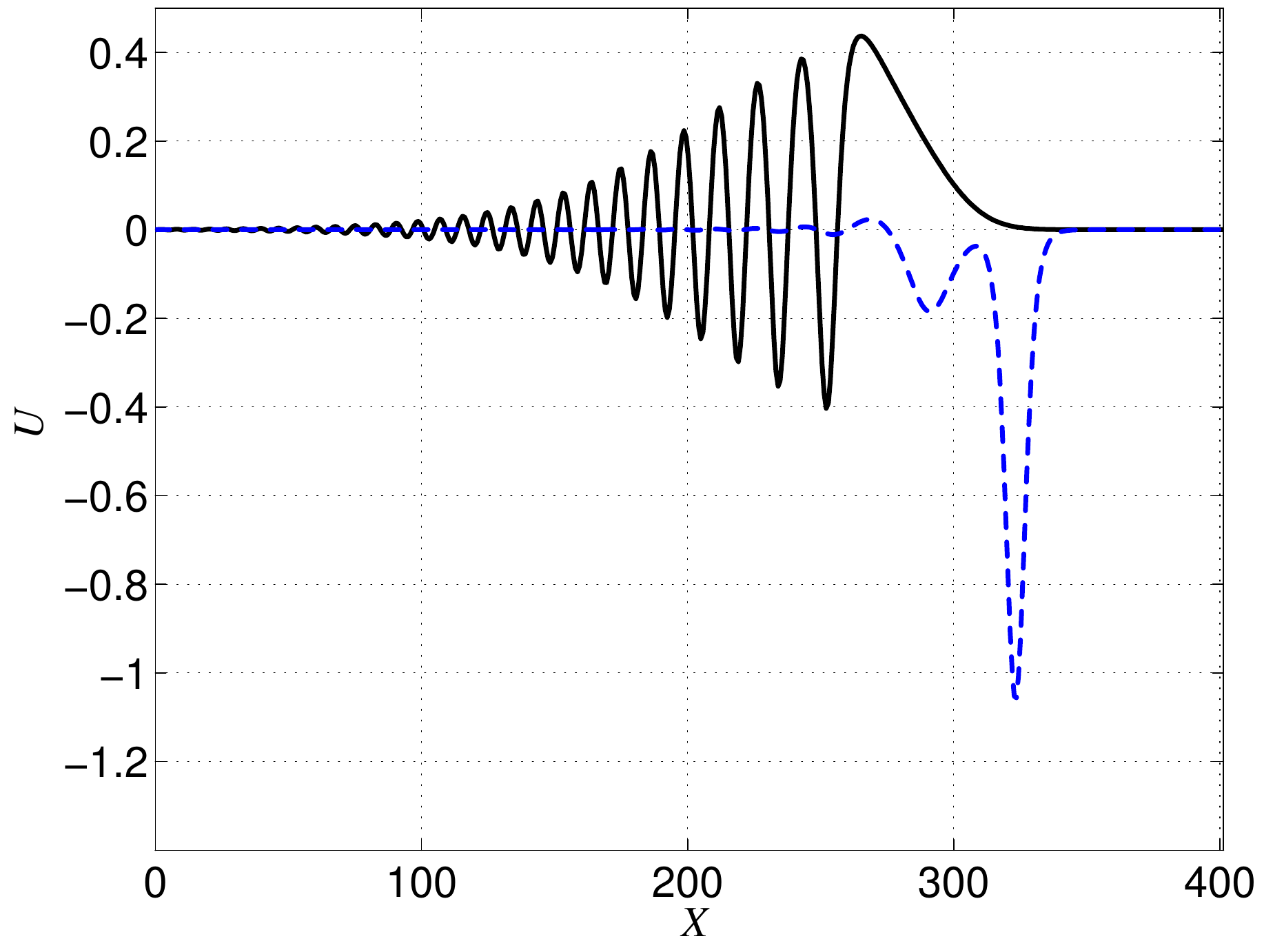}
\includegraphics[width=0.49\textwidth]{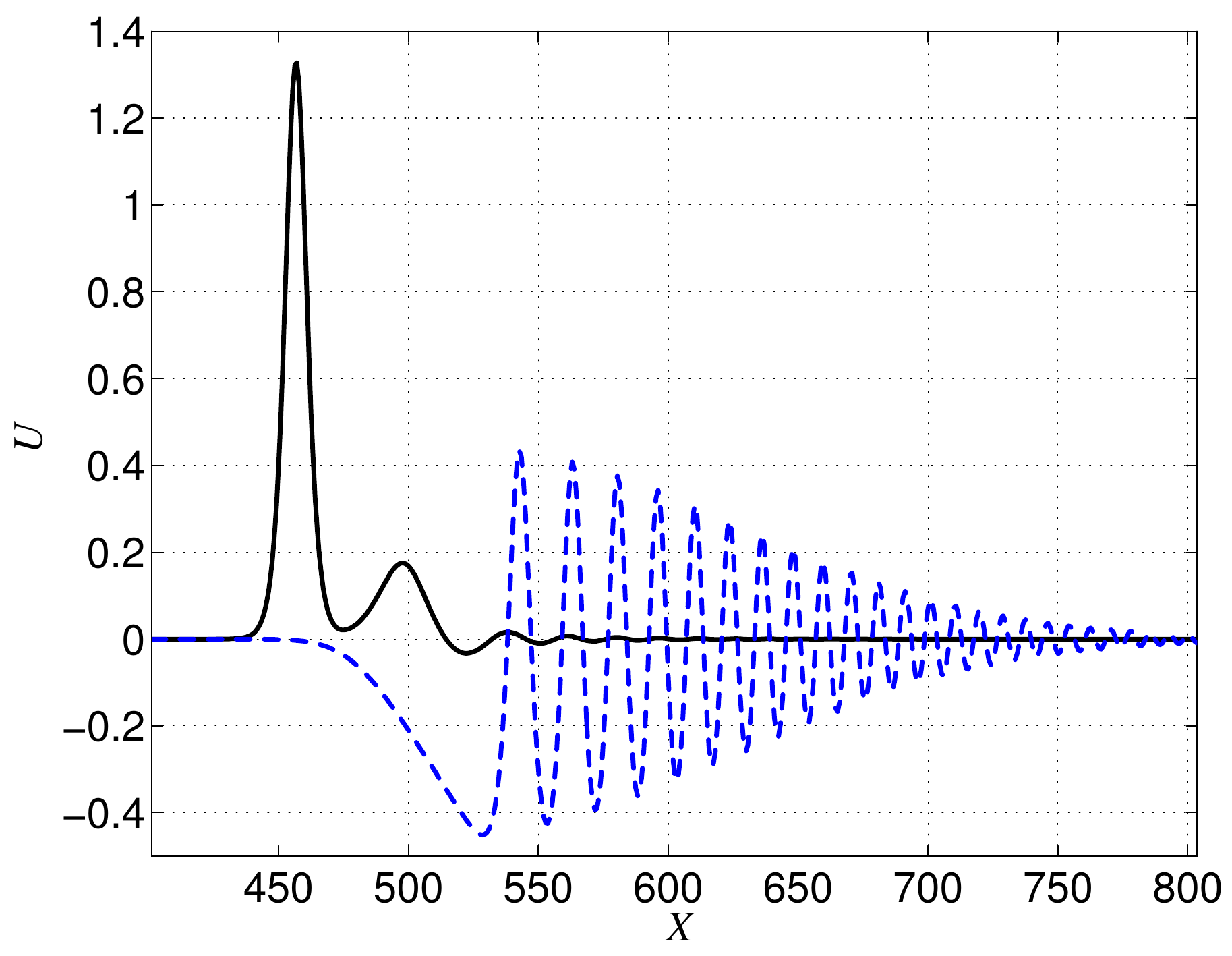}
\caption{Waveprofile plots in the normal (left, $T=1500$) and anomalous (right, $T=1700$) dispersion cases for the positive (solid black line) and negative (blue dashed line) initial condition amplitudes. Waveprofile propagation direction is from left to right. Parameters: $U_o= \pm 1$, $B_o=1/8$, $k=128$, $n=1024$, $c_o=1$, $P=-0.1$, $Q=0.01$, $H_2=0.5$ and $H_1=0.28125$ (normal dispersion), $H_1=0.78125$ (anomalous dispersion).}
%$B,C,D,N$ and $M$.}
\label{Aachenjoon}
%\end{centering}
\end{figure}

\begin{figure}[ht]
%\begin{centering}
%\includegraphics[width=0.95\textwidth]{new3}
\includegraphics[width=0.49\textwidth]{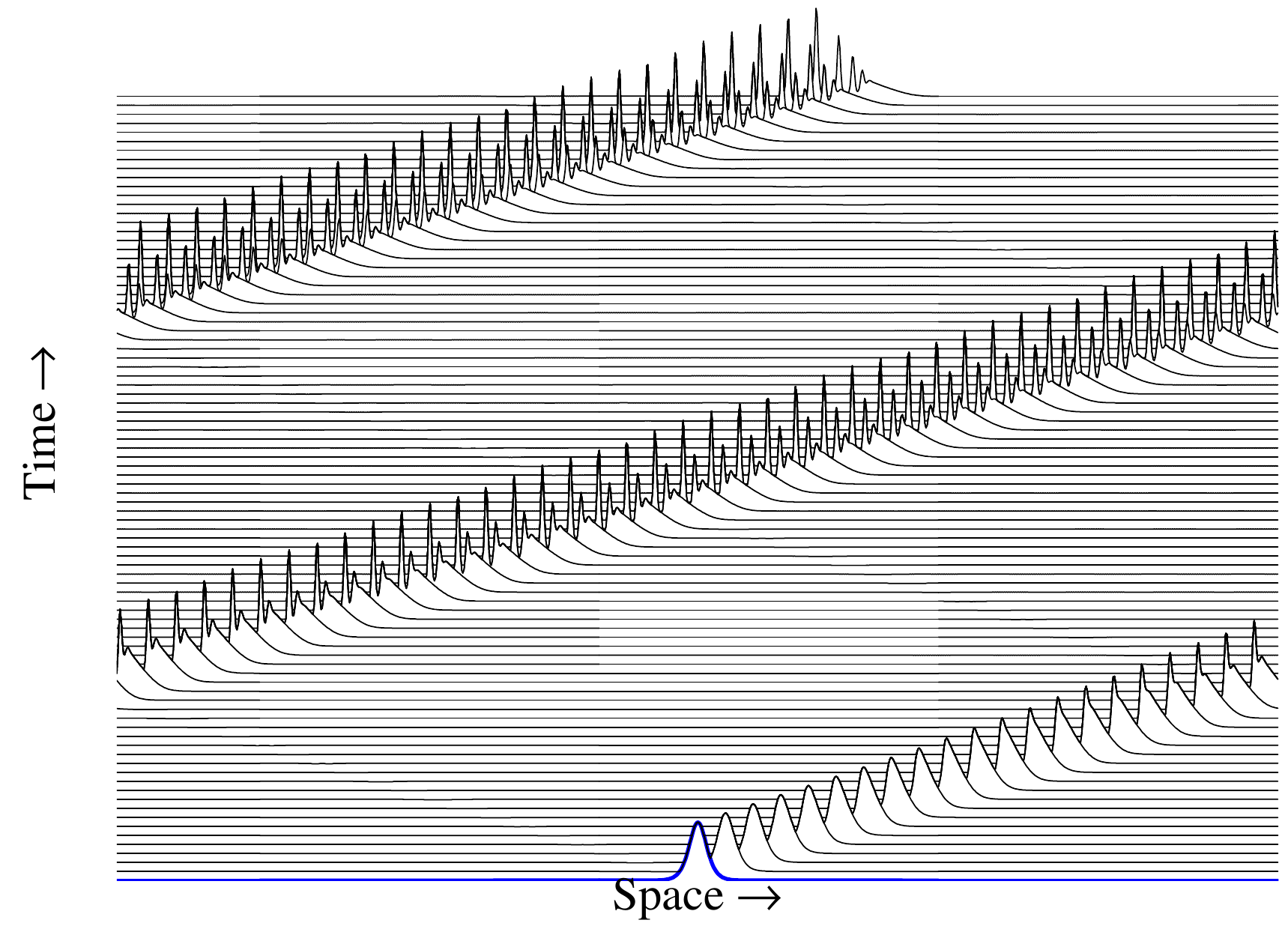}
\includegraphics[width=0.49\textwidth]{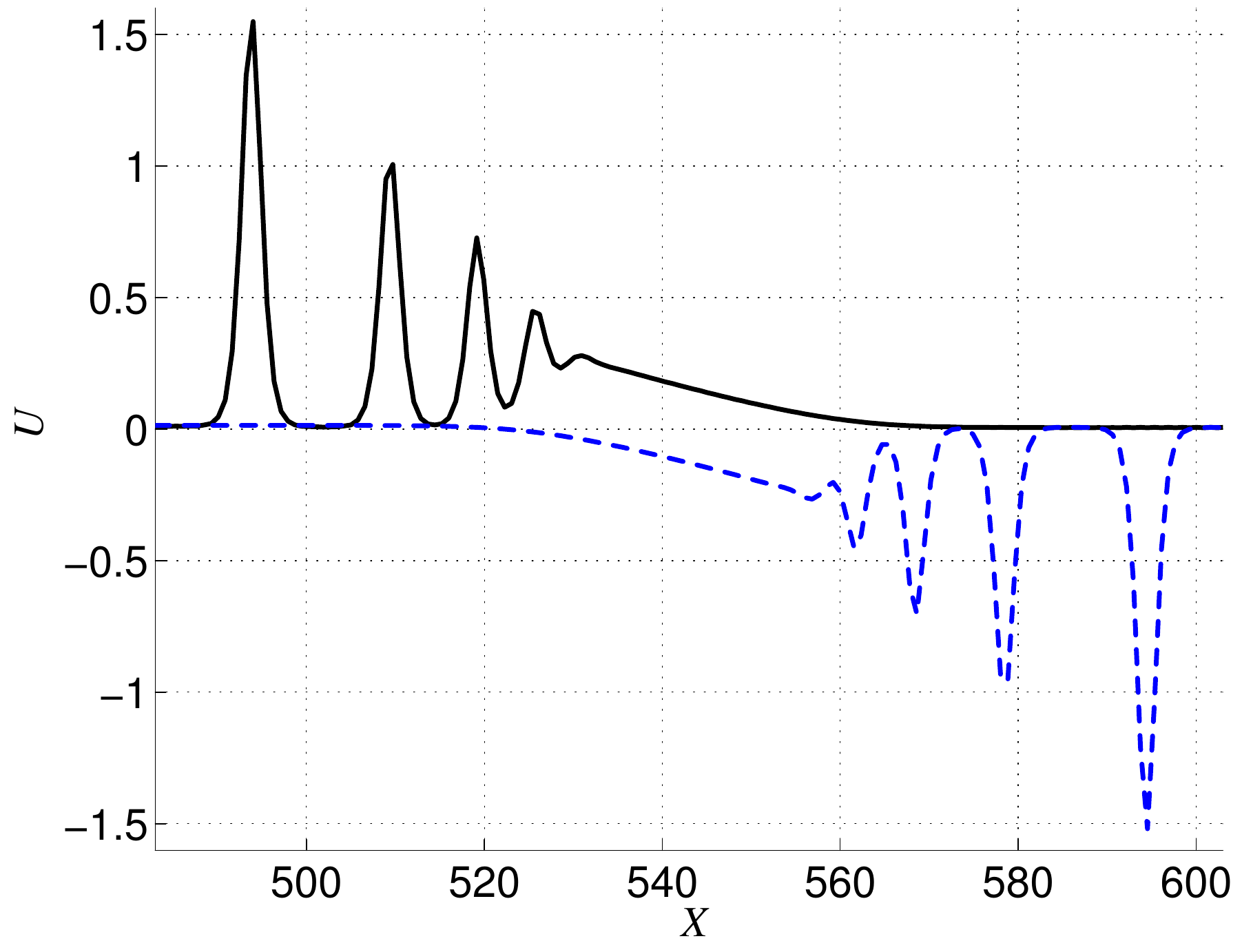}
\caption{The soliton train formation time slice plot in the case of positive initial amplitude (left) and waveprofiles comparison plot (right) at $T=1750$ for the negative (blue dashed line) against positive (black solid line) initial condition amplitude. Lower amplitude solitary waves propagating faster. Direction of propagation from left to right. Parameters: $P=-0.1$, $Q=0.05$, $H_1=0.5$, $H_2=0.5$, $k=128$, $n=1024$, $U_o=\pm 1$, $B_o=1/8$, $c_o=1$, $T=0 \ldots 1750$.}
%$B,C,D,N$ and $M$.}
\label{negpossol}
%\end{centering}
\end{figure}

In addition to the formation of solitary waves a number of different waveprofile regimes exist for the solutions of the governing equations \eqref{EPTdimensionless} depending on the parameters but also on the initial conditions. To name the ones investigated previously:\\ 
(i) solitary waves (single or as a part of solitary wave train, see Figs~\ref{Aachenjoon},\ref{negpossol});\\ 
(ii) Airy or reverse Airy like oscillatory structures (see Fig.~\ref{Aachenjoon});\\ 
(iii) hybrid solution where part of the initial pulse evolves into a train of solitary waves and remainder of the initial pulse forms an oscillatory structure \citep{Peets2015a,Peets2015,Tamm2015}.\\ 
From the viewpoint of nerve pulse propagation the most interesting  one is the solitary wave solution, however, the rest of the solution types can not be ignored either as these might be relevant somehow for either nerve pulse propagation or some kind of pathologies. It should be emphasised that not only the equation parameters are important in determining what kind of solution evolves from the initial excitation but also the character of said initial excitation is important. As an example see Fig.~\ref{Aachenjoon} where some solutions corresponding to the different parameters and initial condition amplitudes are presented. Depending on the dispersion type the initial excitation sign determines if the emergining wave structure is composed of solitary pulses or Airy or reverse-Airy type oscillatory packet under the parameter combinaion used in Fig.~\ref{Aachenjoon}. In essence, a parameter combination which in Fig.~\ref{Aachenjoon} leads to solitary wave train would be of Airy-type if the initial excitation is with opposite sign and vice versa, which previously was of Airy-type solution would be a train of solitary waves. Another interesting phenomenon which must be mentioned is the case where smaller amplitude solitary waves can travel faster than the high amplitude ones as can be seen in Fig.~\ref{negpossol}. Under the suitable parameter combinations it is possible to observe a solutions where both negative and positive amplitude solitary waves can exist simultaneously and if the smaller amplitude solitary waves travel faster then the larger negative amplitude solitary waves travel even faster. 
However, this is not an universal symmetry but needs the right ratio of parameters. The most common solution type seems to be the oscillatory structure with few solitary pulses where some part of the initial pulse energy is sufficient to form one or more solitary pulses and the remainder forms an oscillatory trail either in front or behind (depending of the dispersion type) of the propagating solitary waves. As the system is conservative and we have periodic boundary conditions then the resulting wave profiles keep interacting until after sufficiently long evolution the radiation from not fully elastic interactions causes the solitary waves to merge into the oscillatory structure where they can no longer be detected separately. For all practical purposes the integration time needed for that to happen is so long that before this scenario becomes relevant one would have to question if the decision to not take dissipation into account is still relevant from the viewpoint of physical interpretation of the results.

\begin{figure}[ht]
%\begin{centering}
\includegraphics[width=0.32\textwidth]{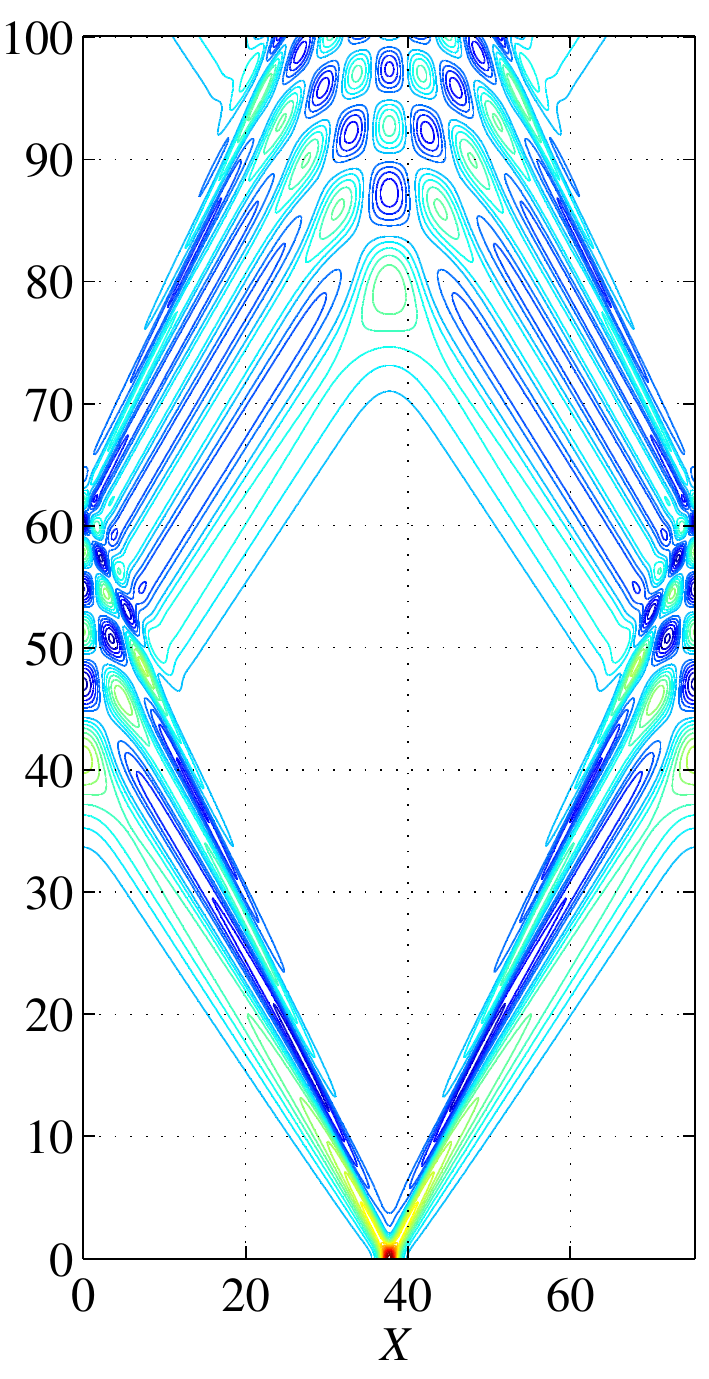}
\includegraphics[width=0.32\textwidth]{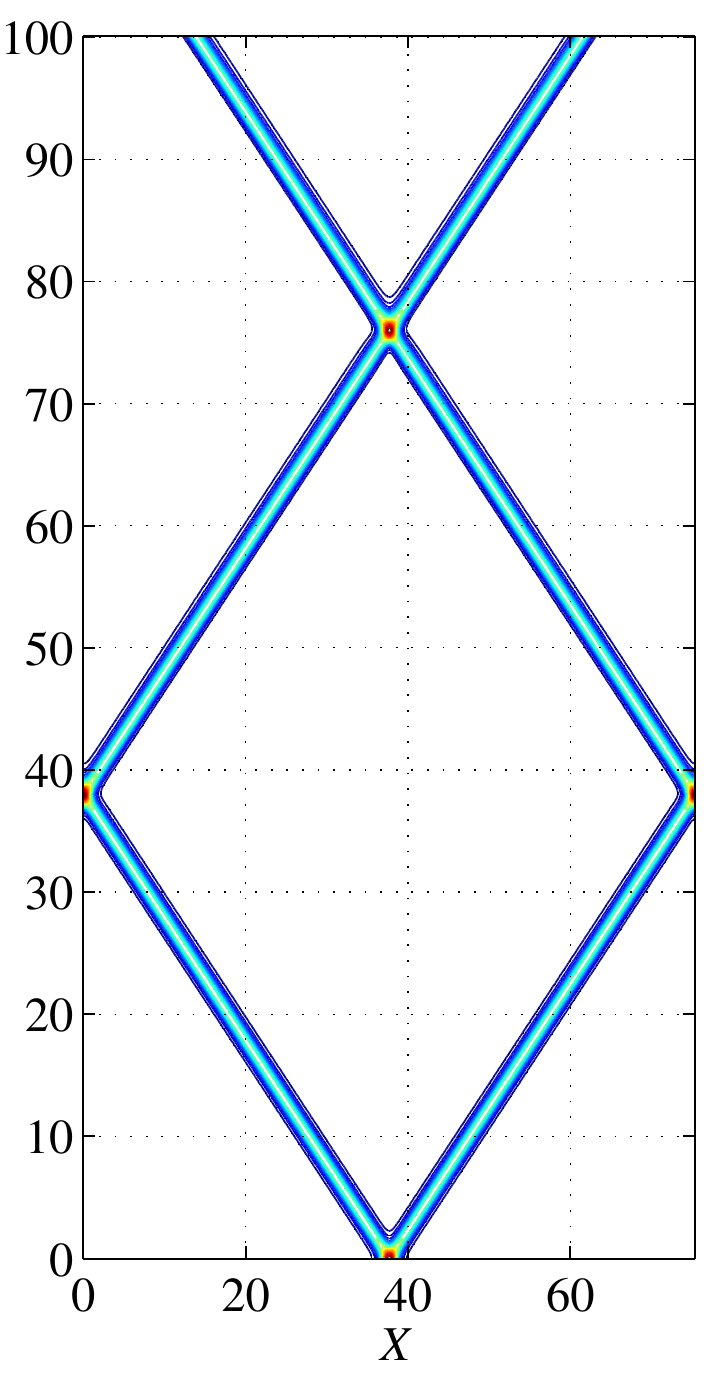}
\includegraphics[width=0.32\textwidth]{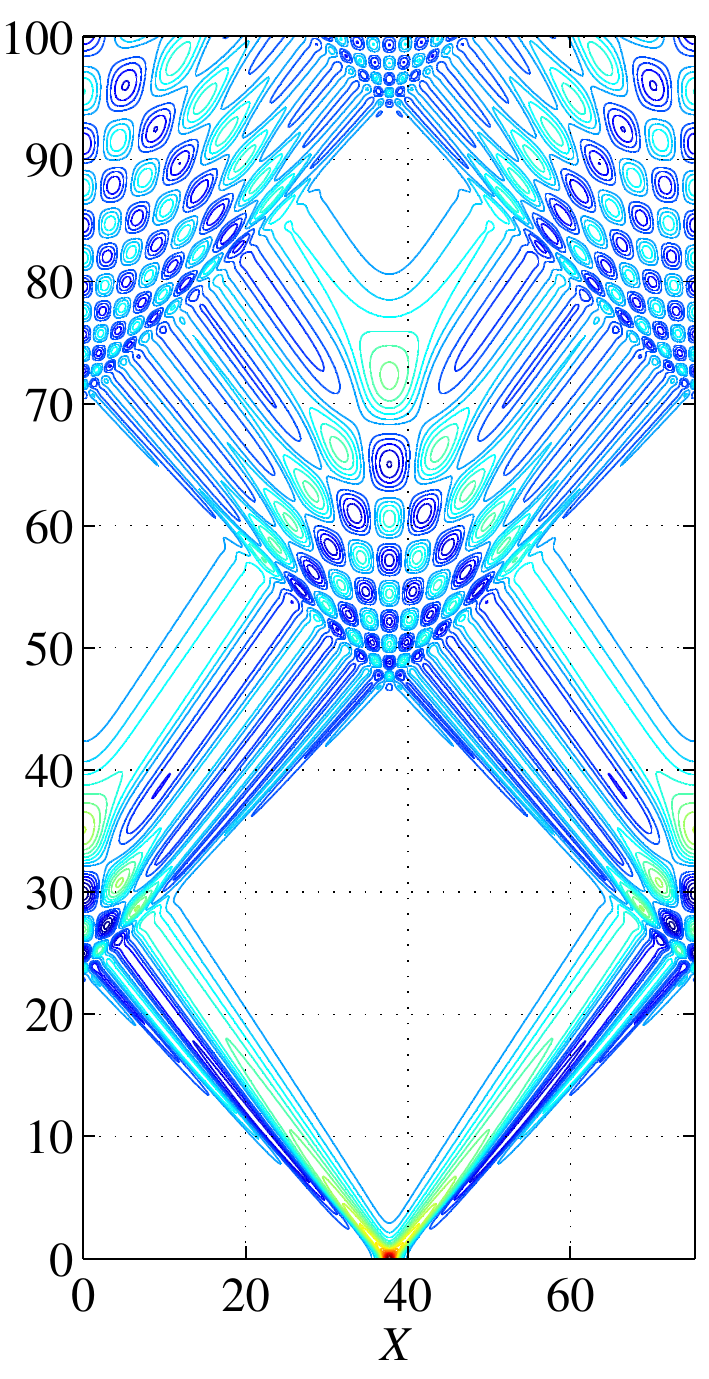}
\caption{The waveprofile contour plots for $P=-0.1$, $Q=-0.1$ and normal dispersion ($H_1=0.3, \; H_2=0.7$, left), balanced dispersion ($H_1=0.5, \; H_2=0.5$, centre) and anomalous dispersion ($H_1=0.7, \; H_2=0.3$, right) cases. Positive initial amplitude. Amplitude isoline interval $0.05$ from $-0.4$ to $+1$, colourmap from blue (negative) to red (positive). Time $T$ on vertical axis.}
%$B,C,D,N$ and $M$.}
\label{trajektoorid1}
%\end{centering}
\end{figure}

\begin{figure}[ht]
%\begin{centering}
\includegraphics[width=0.32\textwidth]{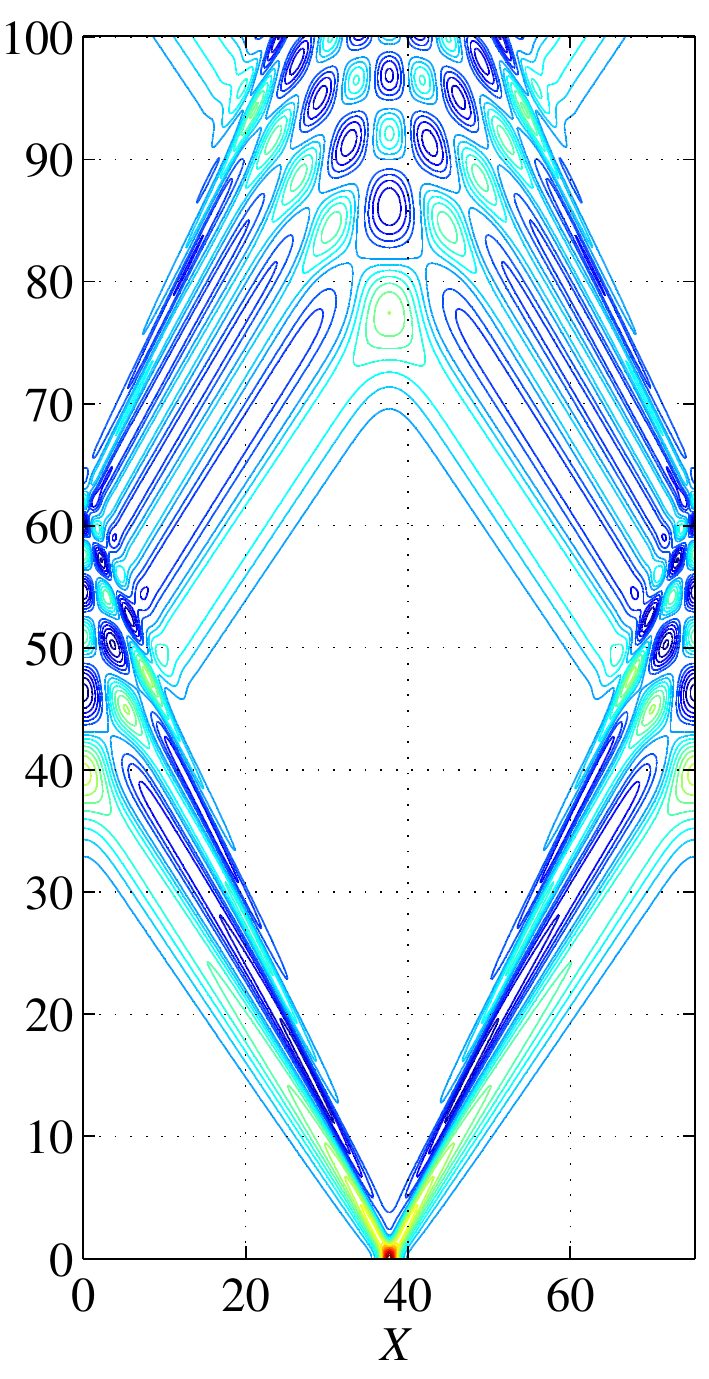}
 \includegraphics[width=0.32\textwidth]{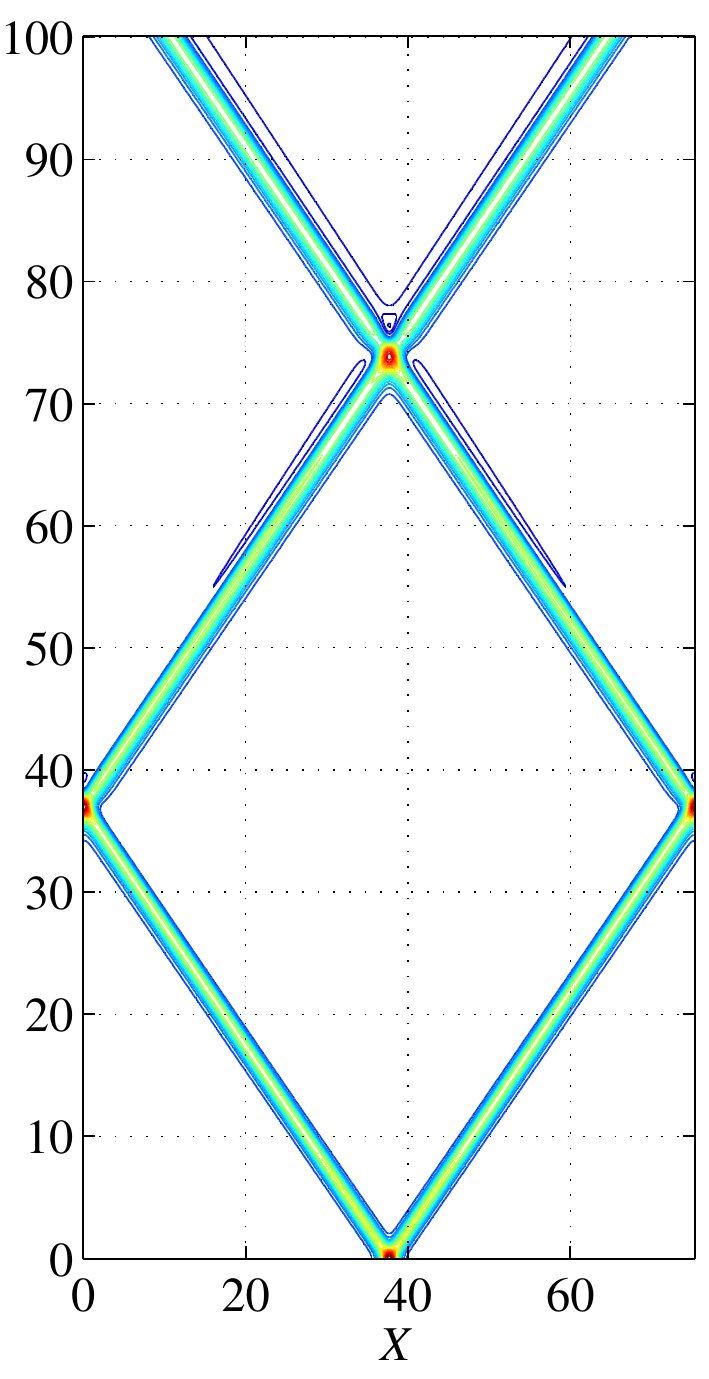}
\includegraphics[width=0.32\textwidth]{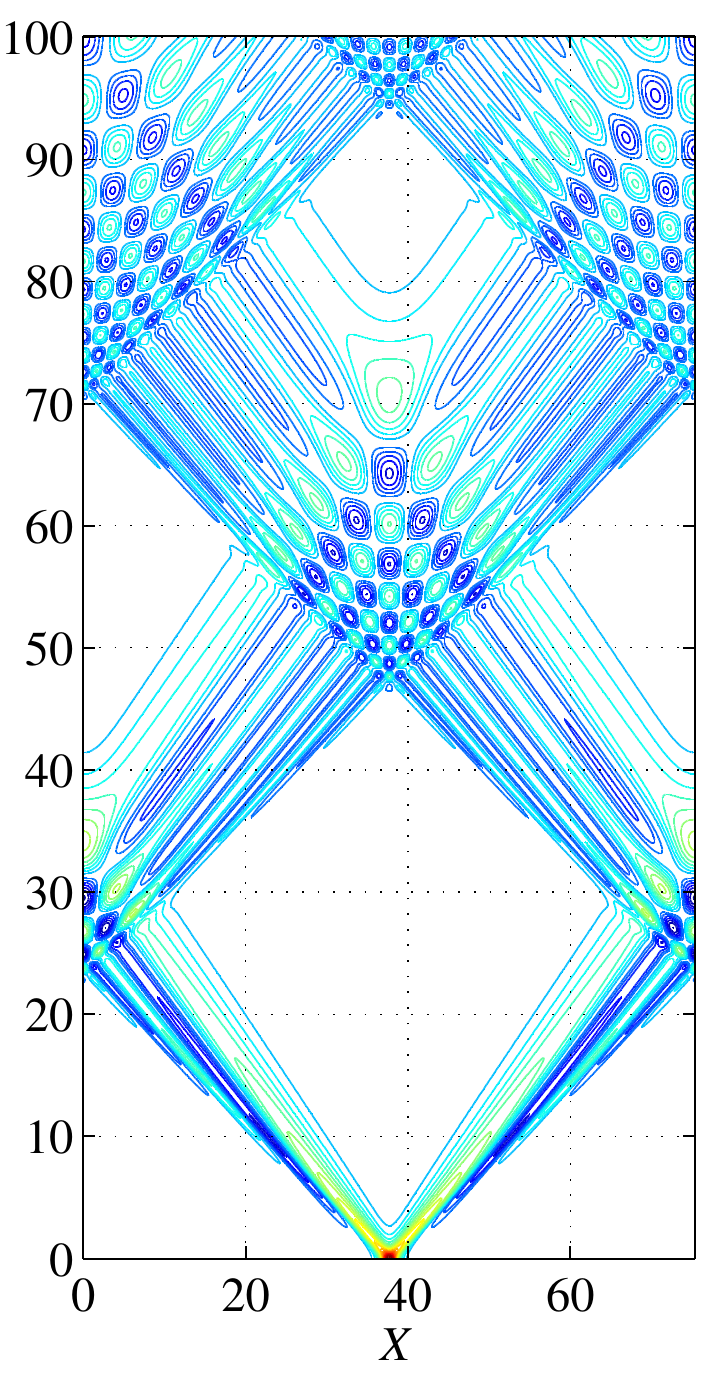}
\caption{The waveprofile contour plots for $P=0.5$, $Q=-0.5$ and normal dispersion ($H_1=0.3, \; H_2=0.7$, left), balanced dispersion ($H_1=0.5, \; H_2=0.5$, centre) and anomalous dispersion ($H_1=0.7, \; H_2=0.3$, right) cases. Positive initial amplitude. Amplitude isoline interval $0.05$ from $-0.4$ to $+1$, colourmap from blue (negative) to red (positive). Time $T$ on vertical axis.}
%$B,C,D,N$ and $M$.}
\label{trajektoorid2}
%\end{centering}
\end{figure}

\begin{figure}[ht]
%\begin{centering}
\includegraphics[width=0.32\textwidth]{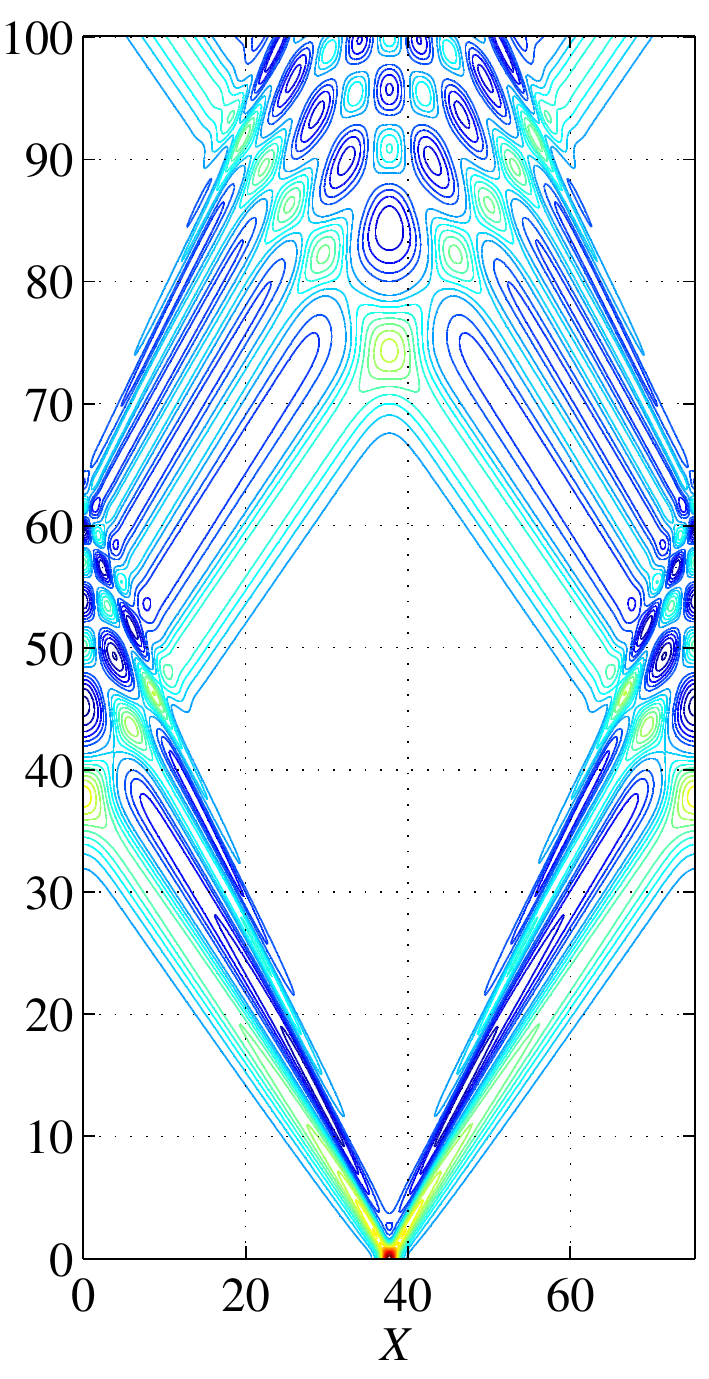}
\includegraphics[width=0.32\textwidth]{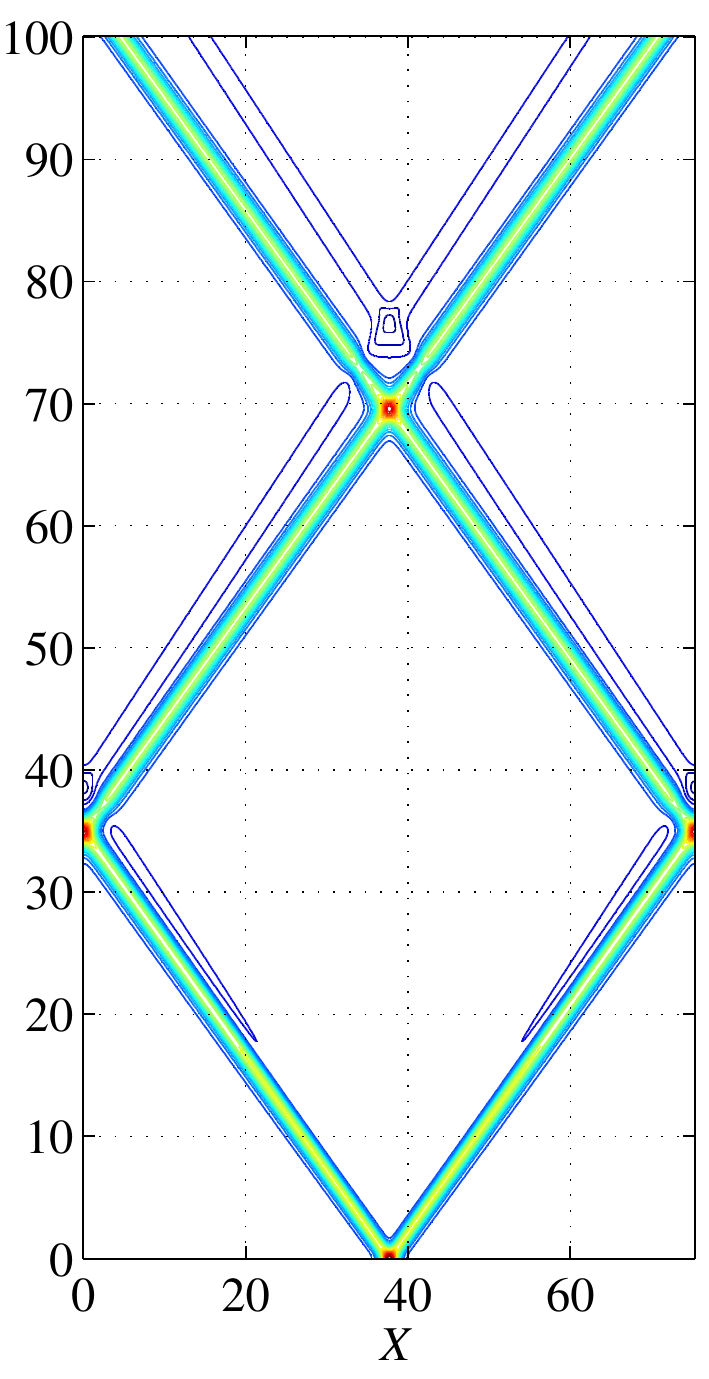}
\includegraphics[width=0.32\textwidth]{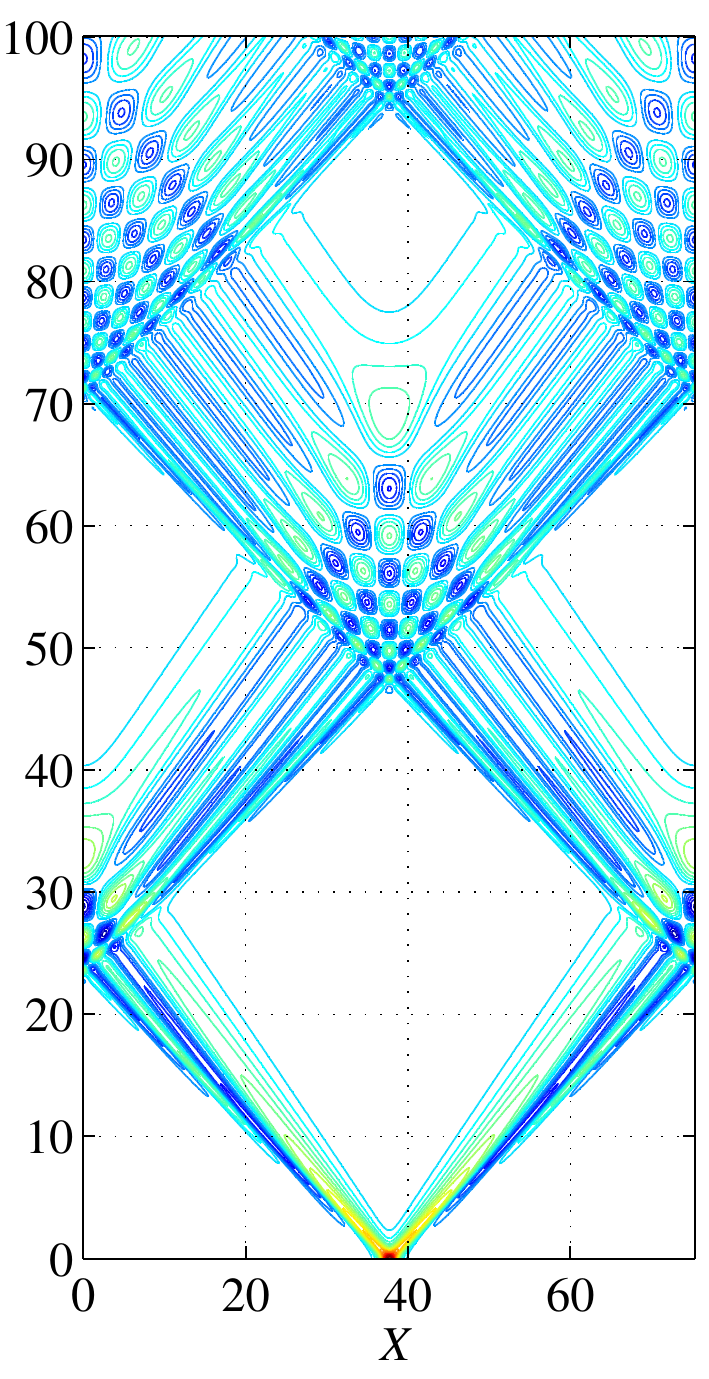}
\caption{The waveprofile contour plots for $P=0.9$, $Q=0.8$ and normal dispersion ($H_1=0.3, \; H_2=0.7$, left), balanced dispersion ($H_1=0.5, \; H_2=0.5$, centre) and anomalous dispersion ($H_1=0.7, \; H_2=0.3$, right) cases. Positive initial amplitude. Amplitude isoline interval $0.05$ from $-0.4$ to $+1$, colourmap from blue (negative) to red (positive). Time $T$ on vertical axis.}
%$B,C,D,N$ and $M$.}
\label{trajektoorid3}
%\end{centering}
\end{figure}

In Figs.~\ref{trajektoorid1}, \ref{trajektoorid2} and \ref{trajektoorid3} one can see selected example countour plots with isoline interval of $0.05$ from $-0.4$ to $+1$ for the amplitude. 
%Two spatial periods are plotted next to each other for making it easier to follow any wave structures that pass through the periodic boundary conditions at $0$ and $24\pi$. 
In addition we are tracking waveprofile peak trajectories by finding the exact local maxima of the wave profiles by making use of the properties of the Fourier transform \citep{Salupere2009} (reconstructing the wave profile from the Fourier spectrum to minimise inaccuracies from using the discrete grid) for finding the exact spatial coordinates of the pulse peaks at each time step for the purposes of finding the waveprofile velocities. Following observations can be highlighted from Figs.~\ref{trajektoorid1}, \ref{trajektoorid2} and \ref{trajektoorid3}: \\
(i) the dispersion parameters have a strong effect on the evolution of the wave profiles -- the main pulse velocities are clearly different depending on the dispersion parameters and in addition the dispersion type determines on which side (relative to the propagation direction) the secondary wave structures emerge from the main pulse. Increasing the parameter $H_1$ increases the main pulse propagation velocity as predicted by dispersion analysis \citep{Peets2015};\\ 
(ii) in the balanced dispersion case the waveprofile remains stable through several interactions even under relatively strong nonlinear parameters (see Fig.~\ref{trajektoorid3}) in addition, it should be emphasised that this is not a classical dispersionless case as both dispersive terms are, in fact, non-zero but are just balanced against each other resulting in a situation where for majority of the frequencies the group and phase speed are equal to each other.\\ 
%It should be added here that in the case of the classical dispersionless case ($H_1=H_2=0$) the wave profiles are not stable and influence of nonlinear terms leads to a formation of singularity in the wave profiles causing the numerical scheme to break down in finite time. The singularity forms in front of the waveprofile (in direction of propagation) if both $P$ and $Q$ are negative regardless if the initial amplitude is positive or negative. If both $P$ and $Q$ are positive then the singularity forms in the front if the initial amplitude is positive and in the back of the pulse if the initial amplitude is negative. If $P$ is negative and $Q$ is positive (as is the case in the original HJ model) then the singularity forms at the back of the pulse regardless if the initial amplitude is positive or negative. 
(iii) The nonlinear parameters $P$ and $Q$ have some influence on the waveprofile propagation velocities (see, for example, Fig.~\ref{trajektoorid1} vs. Fig.~\ref{trajektoorid3}).\\  
In the case of the normal dispersion increasing the nonlinearity leads to the slower propagation velocity for the wave profiles. In the case of the anomalous dispersion the main pulse velocity remains almost the same, however, where the effect is more prominent is the secondary oscillatory structures where in Fig.~\ref{trajektoorid3} the secondary structure starts to separate at approximately $T=2.7$ while in Fig.~\ref{trajektoorid1} the secondary pulses start to separate at approximately $T=2.4$ meaning that in the case of higher nonlinear parameters the secondary structures have propagated at higher velocity. This is in agreement with previous results where it has been demonstrated that due to the uncommon (in the context of Boussinesq type equation) nonlinear terms there can exist parameter combinations where the smaller amplitude solitary waves propagate faster than the higher amplitude ones \citep{Tamm2015}.

%%\newpage
%\begin{figure}[ht]
%%\begin{centering}
%\includegraphics[width=0.47\textwidth]{figure2_4.eps}
%\includegraphics[width=0.47\textwidth]{figure3_4.eps}
%%\includegraphics[width=0.32\textwidth]{traj_nr_3.00_Ao_1.00_P_-0.10_Q_-0.10_H1_0.70_H2_0.30.eps}
%\caption{The formation of trains of solitary waves (see \cite{Tamm2015}). Left -- the case where lower amplitude solitary waves travel faster and right -- the case where the higher amplitude solitary waves travel faster.}
%%$B,C,D,N$ and $M$.}
%\label{soltrains}
%%\end{centering}
%\end{figure}

%\newpage

%\textbf{Have to think a bit more about including the Table and if this is a sensible thing to do in the sense of making it possible to actually show something that is not visible from the figures. It certainly would enable me to write a long text with a lot of numbers [Kert].}

Next, let us take a more detailed look how the the nonlinear and dispersive parameters influence the observable quantities of the wave profiles under the positive and negative initial conditions. 
%For the sake of sanity the width and amplitude of the initial condition are kept constant (other than the sign) and initial phase speed is always zero. 
We observe the speed of the peak of the main pulse.
 %(average between $T=2.5$ and $T=25.5$ based on the coordinates of the peak) after it has separated from the initial condition and the amplitude of the main pulse before first interaction at $T=25.5$. 
We track the coordinates of the peak of the main pulse by reconstructing the waveprofile shape from the full Fourier spectrum at each time step.
 %and finding the $X$ coordinate of maxima to minimise errors from using the discrete grid. In addition we record the absolute maxima and minima of the numerical solutions over the whole integration interval from $T=0$ to $T=100$. 
Parameters $P$ and $Q$ change from $-0.9$ to $+0.9$ with the step size of $0.1$ and for dispersion related parameters $H_1$ and $H_2$ three combinations are recorded -- an normal dispersion case ($H_1=0.3, \, H_2=0.7$), balanced dispersion case ($H_1=H_2=0.5$) and anomalous dispersion case ($H_1=0.7, \, H_2=0.3$). 

(i) $P<0$, $Q<0$:

\emph{Normal dispersion case.} The negative amplitude initial condition results in the faster propagation velocity of the pulse than the positive amplitude initial condition. 
%The minimum amplitude over the integration interval is more or less the same while the main pulse amplitude is higher. 
Increasing $Q$ leads to increase of the observed main pulse velocity and increasing of the main pulse amplitude. On the other hand, increasing the nonlinear parameter $P$ towards zero leads to decrease in main pulse velocity for the negative amplitude initial condition.
The positive initial amplitude case has velocity gains for the main pulse. As far as the amplitude goes, if the initial condition is positive then the main pulse amplitude increases and if negative then decreases. 
The oscillatory structures amplitudes remain roughly the same when nonlinear parameter $P$ changes. 

\emph{Balanced dispersion case.} For the negative initial amplitude case the main pulse velocity is greater than in the case of the positive initial amplitude. 
Increasing the parameter $Q$ leads to increasing main pulse velocity and the main pulse amplitude. Increasing the parameter $P$ towards zero leads to the main pulse velocity moving closer to one (the normalised speed of sound in the present context). If the initial condition is with negative amplitude then increasing parameter $P$ results in the reduction of velocity for the main pulse and if the initial amplitude is positive then the main pulse propagation velocity is increased. The main pulse amplitude is close to half of that of the initial condition and the oscillatory structures are practically non-existing. 

\emph{Anomalous dispersion case.} In the anomalous dispersion case the negative amplitude initial condition results in the faster main pulse velocity in the case of positive initial amplitude than in the case if the initial amplitude is negative. What is different in this case is that increasing $Q$ leads to decrease of the observed main pulse velocity in the case of negative initial condition while in the case of positive amplitude initial condition this leads to increase of the observed main pulse velocity. Increasing the parameter $P$ leads to small increase of the main pulse velocity under both used initial condition signs. 

(ii) $P<0$, $Q>0$:

\emph{Normal dispersion case.} The negative amplitude initial condition leads to a greater main pulse velocity than the positive amplitude initial condition. Under the both initial condition signs decreasing the nonlinear parameter $Q$ (towards the zero) leads to a small decrease of the main pulse velocity. In the case of the negative amplitude initial condition the main pulse amplitude is greater than in the case of the positive amplitude initial condition and the observed oscillations are larger for the case with positive initial amplitude than in the case with the negative initial amplitude. Increasing parameter $P$ leads to decrease in the main pulse velocity in the case of the negative amplitude initial condition while in the case of the positive initial amplitude the main pulse velocity remains the same. Increasing parameter $P$ towards zero leads to marginally greater amplitude for the main pulse in the case of negative amplitude initial condition while in the case of positive initial amplitude the main pulse amplitude is unaffected by the changes in the nonlinear parameter $P$. The oscillatory structure magnitude is unaffected in the normal dispersion case. 

\emph{Balanced dispersion case.} The main pulses propagate with the velocity close to one for the both considered initial condition signs. Decreasing parameter $Q$ leads to a decrease in the observed main pulse velocity and main pulse amplitude under both of the initial condition signs. Increasing parameter $P$ towards the zero leads to main pulse speeds closer to one under both considered initial condition signs. It should be mentioned that the amplitudes of the main pulses are greater than half of the initial pulse height which is due to the nonlinear effects which are combined with relatively weak dispersion under the used parameters combination. 

\emph{Anomalous dispersion case.} The main pulses propagate with velocity greater than one under both of the considered initial condition signs, however, the main pulse amplitudes and associated oscillatory structures are different. Increasing the parameter $Q$ leaves the observed propagation speed the same but decreases the observed main pulse amplitude and leaves the observed oscillatory structures about the same. Increasing the parameter $P$ does not affect the main pulse velocity significantly in the considered dispersion case regardless of the sign of the initial amplitude. However, increasing the nonlinear parameter $P$ decreases the main pulse amplitude and increases the amplitude of the oscillatory structures under the both considered initial condition signs. 

(iii)  $P>0$, $Q<0$:

\emph{Normal dispersion case.} The negative amplitude initial condition leads to a smaller main pulse velocity  than the initial condition with positive amplitude. Regardless of the velocity difference the amplitudes of the pulses are comparable and the same can be observed for the oscillatory tails. Increasing the parameter $Q$ does not affect noticeably the solution corresponding to the negative initial amplitude while in the case of positive initial amplitude  the main pulse velocity and amplitude increase with increasing the parameter $Q$ towards the zero. Increasing the parameter $P$ leads to a small decrease for the main pulse velocity for the negative initial amplitude case and to a increase of the main pulse velocity in the case of positive amplitude initial condition. In addition, increasing nonlinear parameter $P$ decreases the main pulse amplitude in the case of negative initial condition and in the case of positive amplitude initial condition the amplitude of the main pulse is increased when parameter $P$ increases. 

\emph{Balanced dispersion case.} The main pulses tend to propagate at velocity close to one and maintain amplitude which is close to the half of that of the initial condition. Increasing the parameter $Q$ towards zero leads, in general, to the increase of the main pulse velocity and amplitude. Increasing the parameter $P$ leads to decrease of the main pulse velocity and amplitude in the case of the negative initial condition amplitude and to the increase of the main pulse velocity and decrease of the amplitude in the case of positive amplitude initial condition.

\emph{Anomalous dispersion case.} The main pulse propagation velocity is greater than one and the influence of the nonlinear parameters is small as far as the main pulse evolution  is concerned. Increasing the parameter $Q$ leaves the main pulse velocity the same  but reduces the main pulse amplitude by a small amount in the case of the negative initial amplitude. If we have the positive initial amplitude then increasing the parameter $Q$ leads to a reduction of the velocity of the main pulse. Increasing the parameter $P$ leads to a decrease in the main pulse propagation velocity and amplitude under both initial condition signs.

(iv)  $P>0$, $Q>0$:

\emph{Normal dispersion case.} For the considered nonlinear parameters signs the decreasing the parameter $Q$ leads in the case of negative initial amplitude to the decrease of the main pulse and oscillatory structure amplitude while the velocity remains practically the same. In the case of the positive initial condition amplitude there is small decrease in the main pulse velocity if parameter $Q$ decreases while the amplitude and oscillatory structure remain practically the same for the amplitude of the main pulse and for the oscillatory structure. Decreasing the nonlinear parameter $P$ leads to increase of velocity and decrease of the main pulse amplitude and the oscillatory structure amplitude in the case of negative initial condition amplitude. In the case of positive initial amplitude decreasing $P$ leads to the decrease of the main pulse velocity and amplitude and the increase of the oscillatory structure amplitude. 

\emph{Balanced dispersion case.} In the balanced dispersion case the main pulse propagation velocities are close to one and the main pulse amplitudes close to the half of the initial condition amplitude. Decreasing parameter $Q$ in the case of negative initial condition amplitude leads to decrease of the main pulse velocity and the amplitude of the oscillatory structure. In the case of positive initial condition amplitude decreasing parameter $Q$ leads to marginally smaller main pulse amplitude and marginally larger oscillatory structure amplitude with minor drop also in the main pulse propagation velocity. Decreasing the parameter $P$ leads in the case of negative initial condition amplitude to a greater propagation velocity of the main pulse as well as to a increased main pulse amplitude and decrease of the oscillatory structure amplitude close to zero. Decreasing the parameter $P$ in the case of positive initial amplitude leads to a decrease of the main pulse velocity while the amplitude of the pulse is increased and the oscillatory structure amplitude is suppressed.

\emph{Anomalous dispersion case.} Decreasing the parameter $Q$ leads to a marginal decrease of the main pulse and oscillatory structure amplitude if the initial condition is with negative amplitude and to a increase of the main pulse velocity and amplitude and to a decrease of the oscillatory structure amplitude if the initial condition has a positive amplitude. Decreasing the parameter $P$ leads to a small decrease in the main pulse velocity for the amplitude and increase of the oscillatory structure amplitude if the initial condition is with negative amplitude. If the initial condition has a positive amplitude then decreasing the parameter $P$ leads to a increase in the main pulse velocity and amplitude and to a decrease of the oscillatory structure amplitude.

\section{Interaction of solitons}

\begin{figure}[h]
%\begin{centering}
\includegraphics[width=0.95\textwidth]{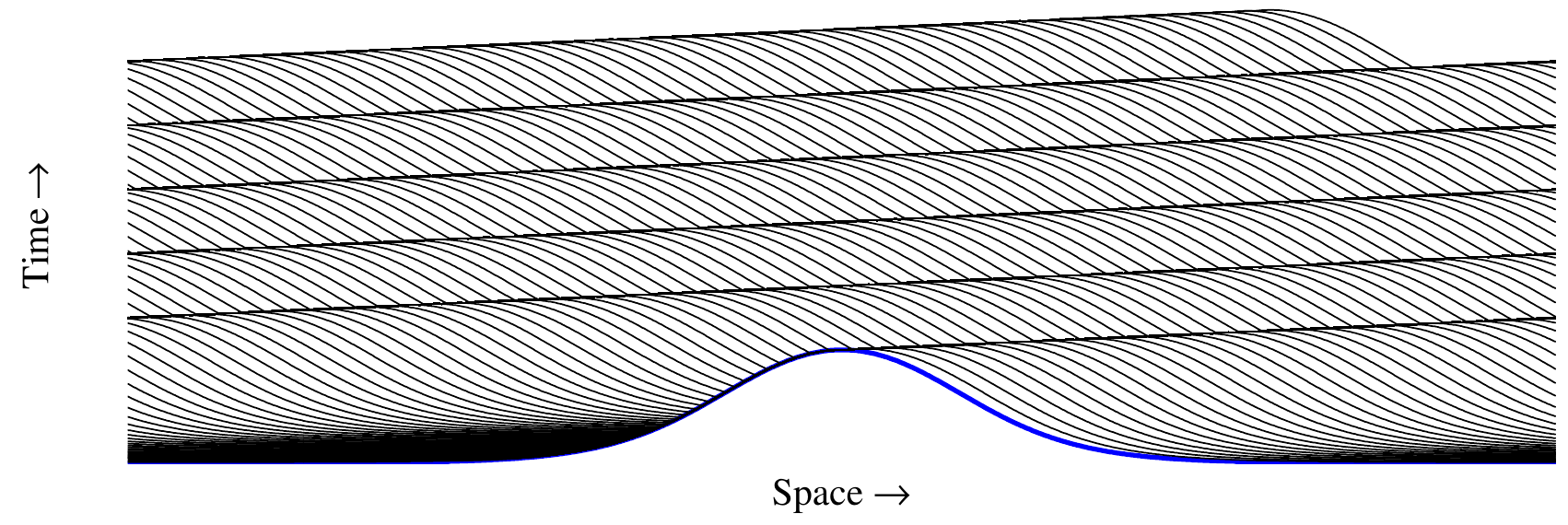}
\includegraphics[width=0.95\textwidth]{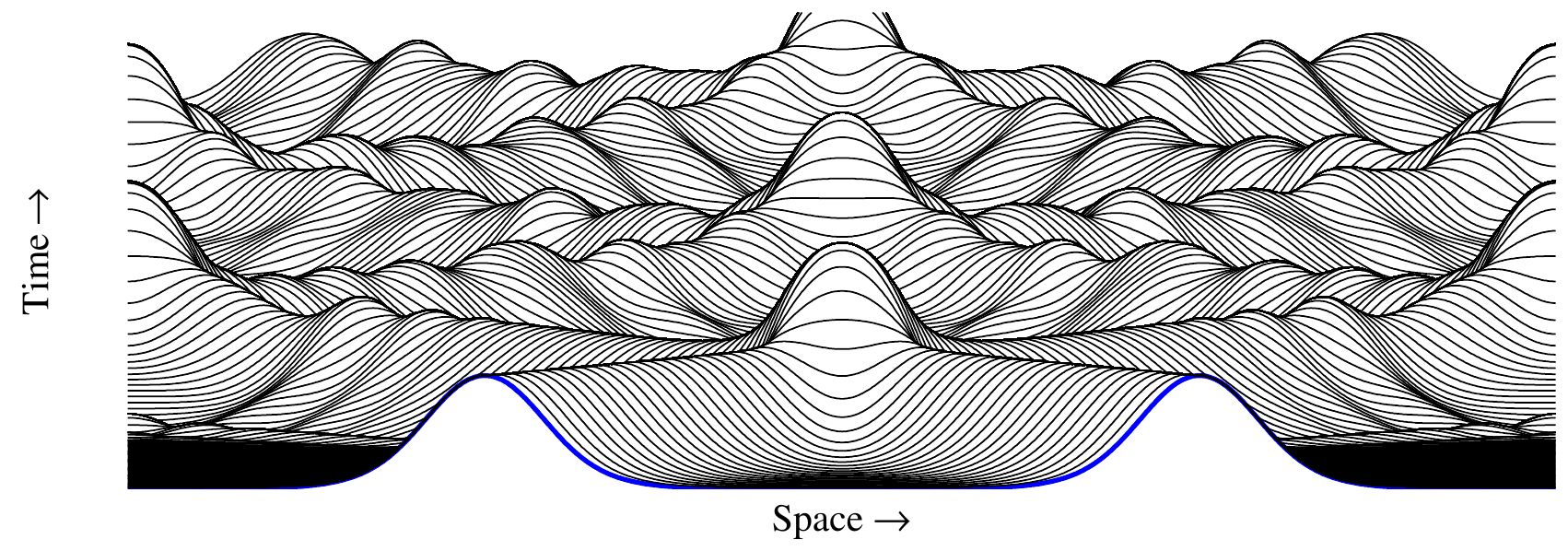}
\caption{The timeslice plots for the case $H_2=0$ Eq.~\eqref{HJ}. The solitary wave solution (top) and interaction of solitary waves (bottom).}
%$B,C,D,N$ and $M$.}
\label{HJwpildid}
%\end{centering}
\end{figure}

In general, a soliton can be described as a stable particle-like state of a nonlinear system \citep{SaNWE1982}. Another way of describing the phenomenon we call soliton is through its properties. A soliton is a wave in the nonlinear environment that (1) has a stable form, (2) is localized in space and (3) restores its speed and structure after interaction with another soliton \citep{Drazin1989,engelbrecht1995}. Solitons emerge when there is a balance in the system between dispersive and nonlinear effects. In essence it can be said that solitons are nonlinear waves that behave between interactions like linear waves. A solitary wave is usually a wave in the nonlinear environment where all the key properties of solitons are not strictly fulfilled. For example, if the interaction between two waves is not entirely elastic (or it is not possible to observe the interaction) or if the form of the wave is not sufficiently stable in time, then the wave is often called a solitary wave to distinguish it from the soliton.

\begin{figure}[ht]
%\begin{centering}
\includegraphics[width=0.95\textwidth]{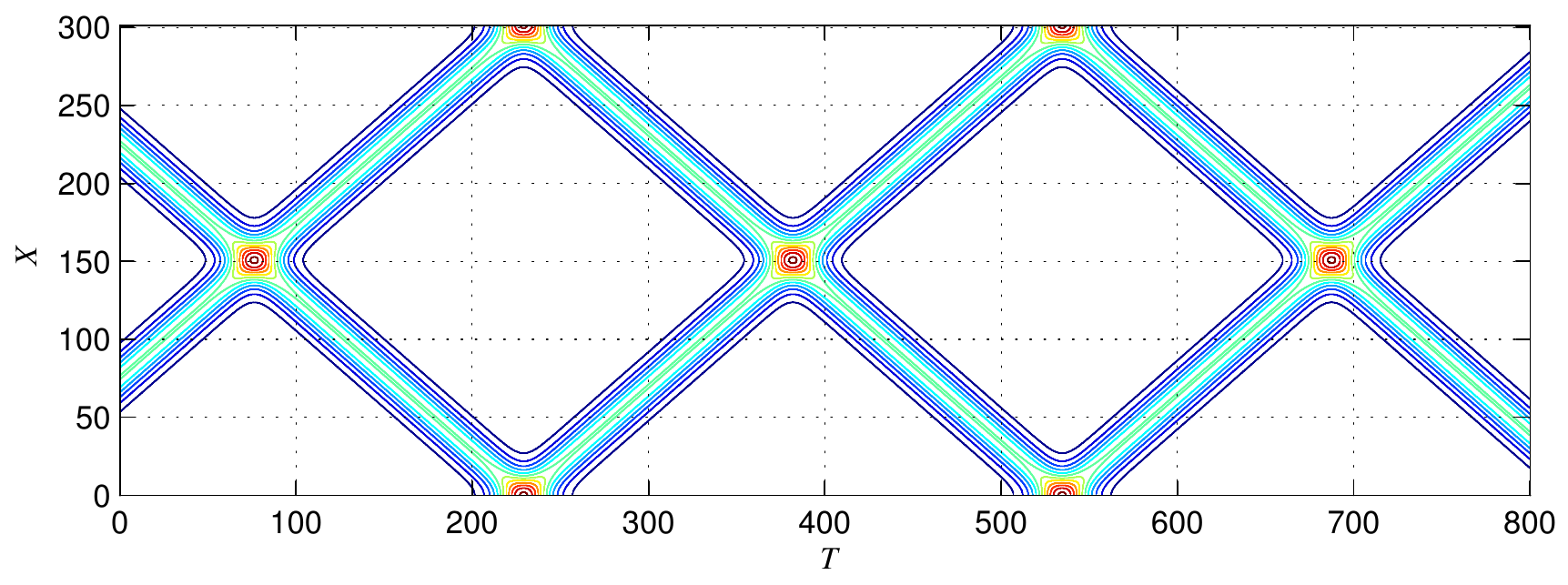}
\includegraphics[width=0.95\textwidth]{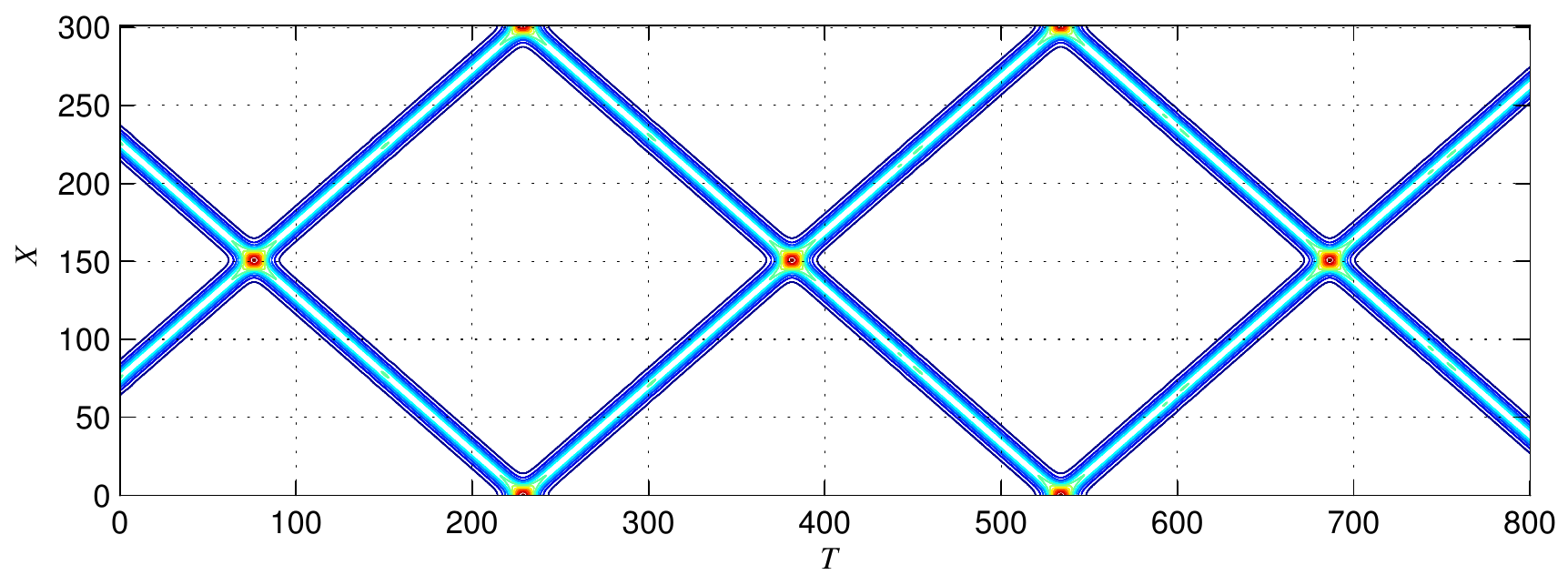}
\caption{Contour plots of interactions of solitonic solutions. Parameters $c=\pm 0.99$, $P=-10$, $Q=40$, $H_1=1$, $H_2=0$ (top) and $H_2=0.75$ (bottom). }
%$B,C,D,N$ and $M$.}
\label{Interactioncontour}
%\end{centering}
\end{figure}

In Fig.~\ref{HJwpildid} one can see the HJ model \eqref{HJ} solitary wave prpagation (top) and interaction (bottom). The parameters are the same as in Section 3 except $H_2=0$. From Fig.~\ref{HJwpildid} it is clear that while the single HJ pulse is stable it is a solitary wave, not a soliton, because the interaction with another such wave is not elastic as there is significant radiation even during the first interaction event and the shape of the waveprofile is not properly restored after the interaction. However, it should be noted that the parameter combinations can exist where the solitary wave solutions can be relatively stable with almost no radiation.

\begin{figure}[ht]
%\begin{centering}
\includegraphics[width=0.475\textwidth]{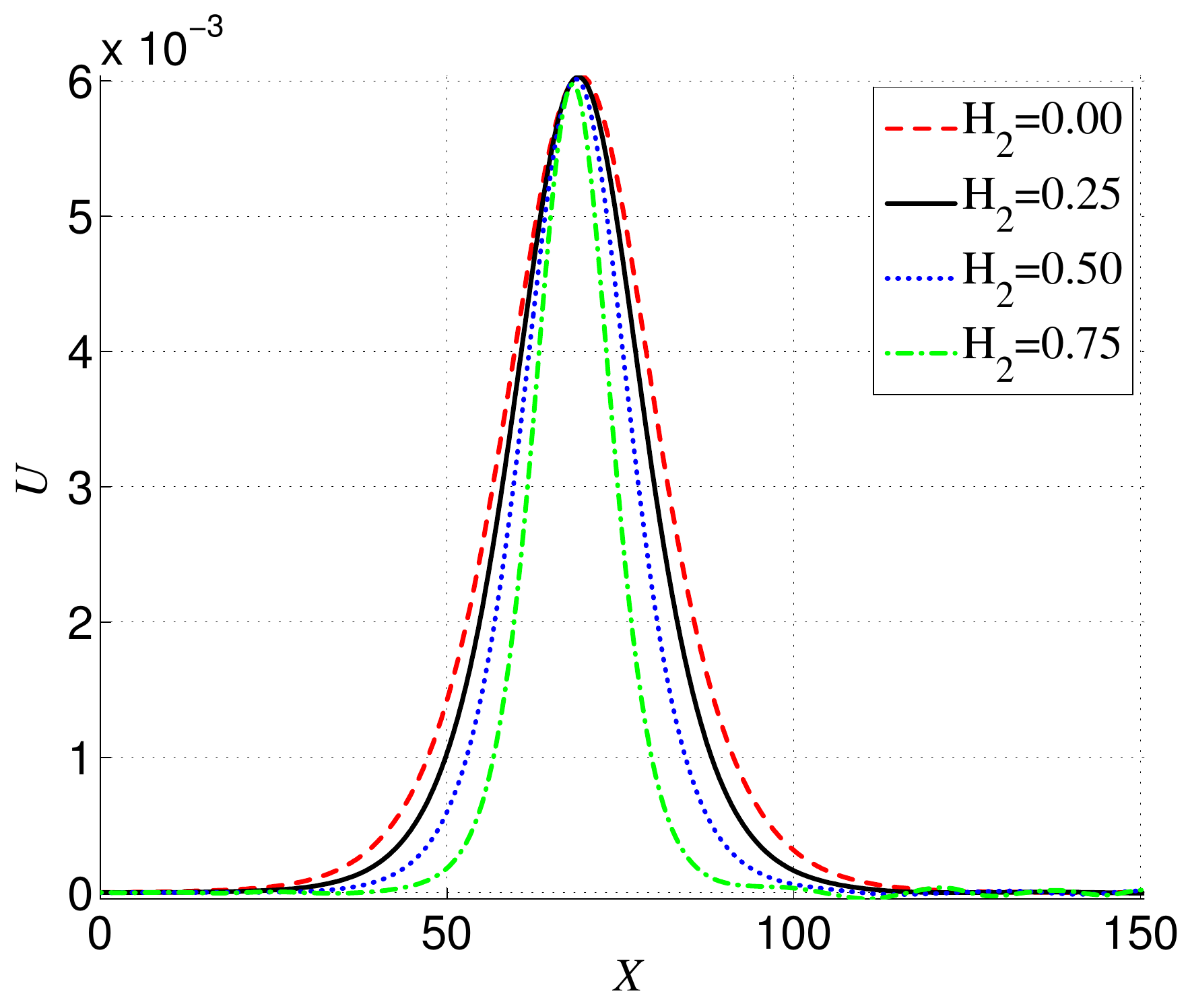}
\includegraphics[width=0.475\textwidth]{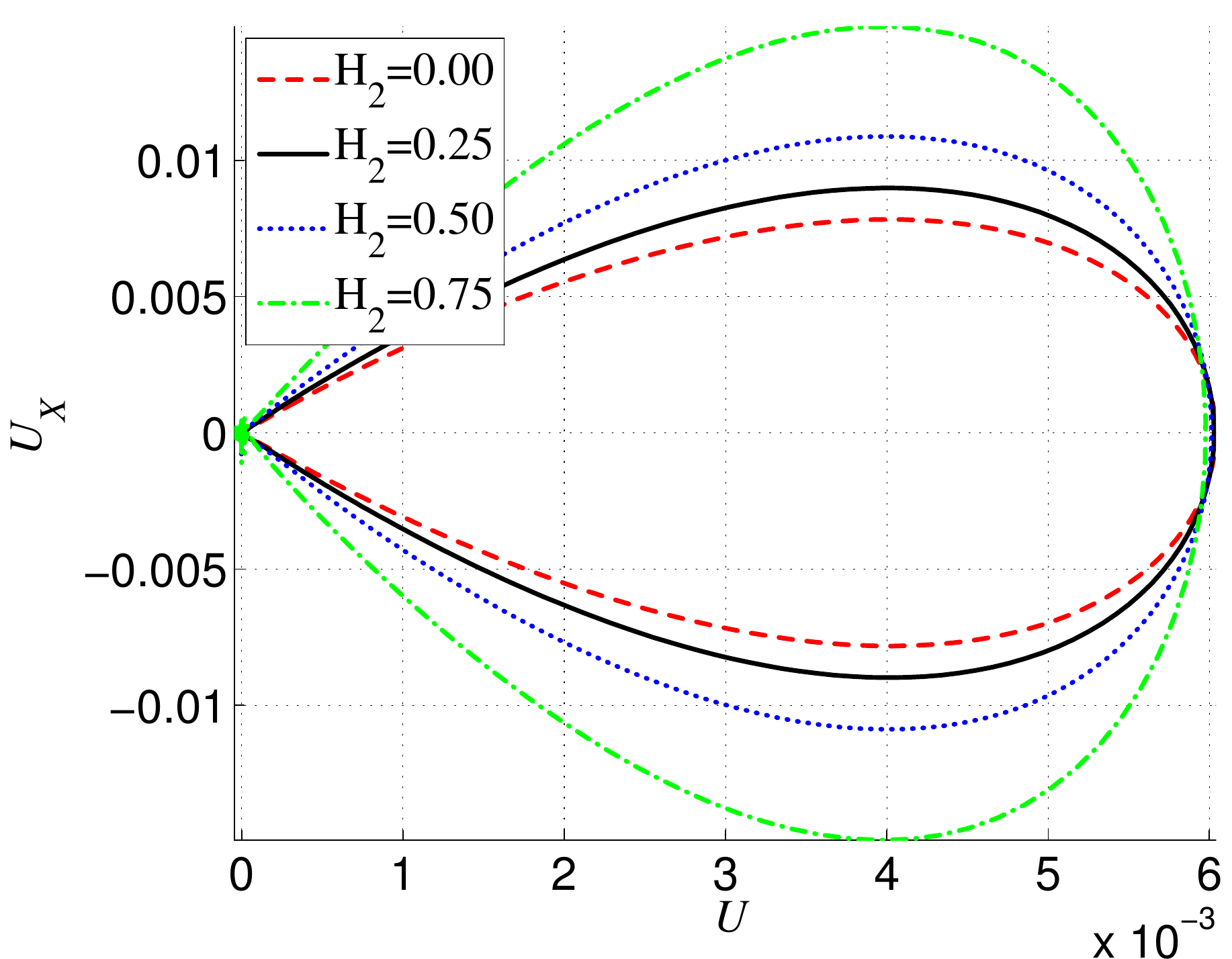}
\caption{Waveprofile plots (left) and corresponding phase plots (right) at $T=770$ after five interactions. Parameters $c=\pm 0.99$, $P=-10$, $Q=40$, $H_1=1$. Only a waveprofile propagating to the left is shown.}
%$B,C,D,N$ and $M$.}
\label{InteractionWprofiles}
%\end{centering}
\end{figure}

The interaction of single solitary waves depends on the parameters of he model and consequently, on the dispersion type (normal, anomalous). We start with the following set
%Let us reduce the dispersive parameters and switch a dispersion type from normal (as it is in Fig.~\ref{HJwpildid}) to anomalous. Picking the parameters as 
$c=\pm 0.99$, $P=-10$, $Q=40$, $H_1=1$, $H_2=0$, $H_2=0.25$, $H_2=0.50$ and $H_2=0.75$.
% allows us to observe a situation where the interactions have almost no radiation (Figs \ref{Interactioncontour} and \ref{InteractionWprofiles}). 
In Fig.~\ref{Interactioncontour} one can see the interactions if the parameter $H_2=0$ and in the bottom if $H_2=0.75$ -- the interactions are remarkably similar and non disruptive with the main difference being that the solitonic waveprofiles are more localized if $H_2 \neq 0$. In this case the interactions have almos no radiation (negligible radiation two orders of magnitude smaller than the main pulse amplitude at $\approx 10^{-5}$). Amplitude isolines are separated by $0.001$ from 0.001 to 0.013 in Fig.~\ref{Interactioncontour}.
%There is still minor radiation but it is two orders of magintude smaller than the main pulse amplitude ($\approx 10^{-5}$) and as such not visible with naked eye on the Figure. 

\begin{figure}[ht]
%\begin{centering}
\includegraphics[width=0.95\textwidth]{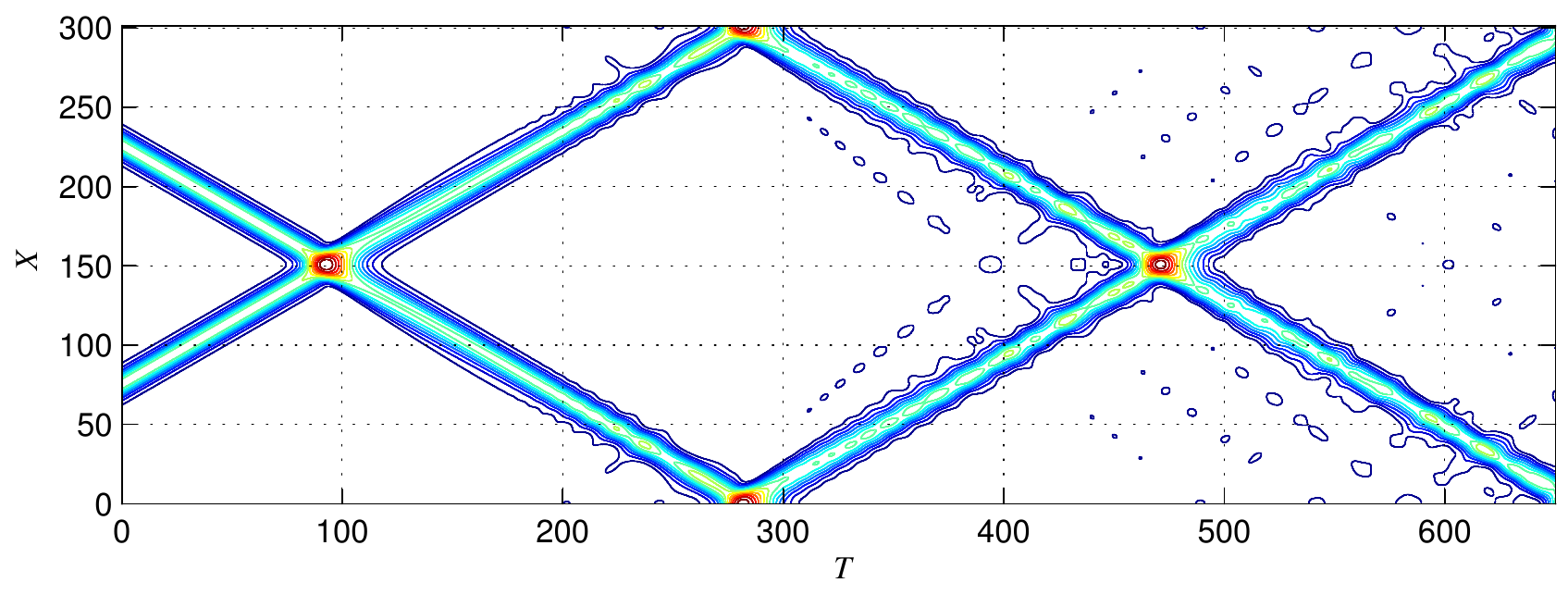}
\includegraphics[width=0.95\textwidth]{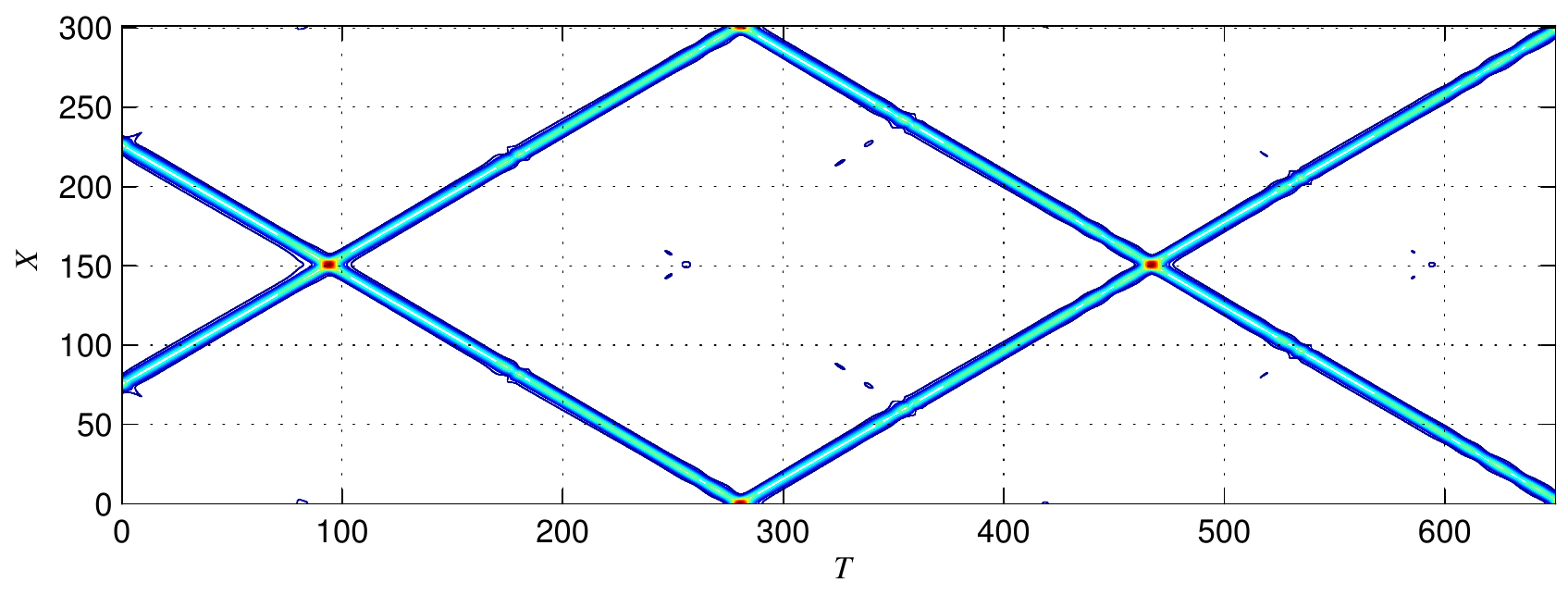}
\caption{Contour plots of interactions of solitonic solutions. Parameters $c=\pm 0.8$, $P=-10$, $Q=40$, $H_1=4$, $H_2=0$ (top) and $H_2=5$ (bottom). }
%$B,C,D,N$ and $M$.}
\label{Interactioncontour2}
%\end{centering}
\end{figure}

In Fig.~\ref{InteractionWprofiles} waveprofiles and corresponding phase plots after the five interaction events ($T>700$) are presented. The solitonic waveprofiles corresponding to higher values of $H_2$ are more localised as expected (the waveprofile in Fig.~\ref{InteractionWprofiles} is propagating to the left).
% however, in addition to that one can note that the waveprofiles corresponding to higher values of $H_2$ have propagated a bit further after five interactions . 
The small distortions to the waveprofiles are easier to spot in phase plots (right), in particular the small radiation close to zero which is two orders of magnitude smaller, as noted, than the main pulse under the used parameter combination.

Let us return to a parameter set presented in Section 3 for the analytical solution. It turns out that there is also a scenario possible where the solitonic solutions with the additional dispersive term are more stable through interactions than the solitonic solutions if parameter $H_2=0$. In Fig.~\ref{Interactioncontour2} the case $H_2=0$ is presented at the top and the case $H_2=5$ in the bottom. The amplitude isolines are separated by 0.02 from 0.02 to 0.3. It is clear that under the parameter set used in Figs \ref{HJwpildid} and \ref{Interactioncontour2} the solitonic waves corresponding to $H_2=0$  have greater amount of radiation than the case $H_2=5$ which is relatively stable in comparison throughout interactions. Neither of the cases can be considered solitons in the strict mathematical case \citep{Drazin1989} as in both cases there is significant enough radiation after only three interactions. While unrelated to the mechanics of soliton interactions it is interesting to remark  that the used numerical algorithm performs approximately three times faster if $H_2 \neq 0$.

In addition it should be noted the the parameter set used for the systematic analysis in Section 4 has non-neglible amounts of radiation during the interactions as well.  However, this is not easy to see in Figs.~\ref{trajektoorid1}, \ref{trajektoorid2} and \ref{trajektoorid3} at the given scale. For that parameter set the amplitude loss due to the interaction events is less than 10\% over dozen interaction events.

\section{Final remarks}

The systematic analysis of solutions to the special Boussinesq-type equation with the displacement-dependent nonlinearities has revealed several interesting phenomena. The analysis is focused on Eq.~\eqref{EPT} (or its dimensionless form \eqref{EPTdimensionless}) which is the improved Heimburg-Jackson model for describing the longitudinal wave process in biomembranes. Like every wave equation it describes the process generated by initial and/or boundary conditions expressed in terms of the dependent variable. Here the variable under consideration is the change of the density in the longitudinal direction. In terms of this variable the existence of solitary solutions is demonstrated, the emergence of trains of solitary pulses is shown and the properties of emergence analysed, and the interaction of single solitary waves and trains studied. The governing nonlinear wave equation is actually a novel mathematical model compared to the conventional models in continuum theory where the nonlinearities as a rule are of the deformation-dependent.

The analysis can be summarised with following conclusions:
\begin{itemize}
	\item The improved model (Eqs \eqref{EPT}, \eqref{EPTdimensionless}) removes the discrepancy that at higher frequencies the velocities are unbounded (see Fig.~\ref{Dispfig});
 \item The additional dispersive term $u_{xxtt}$ with the coefficient $h_2$ (or $H_2$ in the dimensionless form) in addition to the ad hoc dispersive term $u_{xxxx}$ \citep{Heimburg2005} describes actually the influence of the inertiality of the microconstituents (lipids) of the biomembrane. This corresponds to the understandings of continuum mechanics of microstructured solids (Mindlin, 1964) and is demonstrated also experimentally \citep{MaurinPHD}. This term regulates the width of the solitary pulse (see Fig. \ref{solitonwidth}) and such an effect can be used for determining the value of $h_2$ from experiments. It also determines how fast the transition from the low frequency speeds to the high frequency speeds occurs (see Fig.~\ref{Dispfig}); 
 \item The fourth-order pseudopotential \eqref{Poly} involves several solution types of solitary waves and under certain conditions ($Q > 0$, $H_2 >> H_1$) an oscillatory solution exists (see Fig.~\ref{Fig4});
 \item Soliton trains can be emerged from an arbitrary initial condition. These results were obtained by numerical simulation by using the pseudospectral method \citep{Salupere2009}. Depending on the signs of coefficients $Q$ and $P$, the nonlinear effects start to influence the emergence either from the front or from the back of the propagating pulse (see Fig~\ref{negpossol}). For the case of a biomembrane one has $Q <0$, $P > 0$ and the train emerging from a positive input starts with smaller solitons which travel faster than the bigger ones. This is different from the conventional case of nonlinear evolution equations (the KdV equation, for example). In the case of a negative input, the train is headed by bigger solitons which travel faster (see Fig.~\ref{negpossol}). It has been shown that there are several wave types possible: solitary waves (Fig.~\ref{Aachenjoon}), oscillatory (Airy-type) waves (Fig.~\ref{Aachenjoon}), and hybrid solutions.
 \item The interaction of solitary waves is not fully elastic (see Figs \ref{Interactioncontour}, \ref{Interactioncontour2}) which shows that these solitary waves are not solitons in the strict sense \citep{Drazin1989}. However, like in other Boussinesq-type equations \citep{Christov2007,Engelbrecht2011}, the radiation effects accompanying every interaction start cumulating rather slowly and the interacting solitons keep their shape for a rather long time. It gives the ground to call emerging solitary waves modelled by Eq.~\eqref{EPT} (or Eq.~\eqref{EPTdimensionless}) solitons like it is done in other physical cases \citep{Maugin2011}.
	\end{itemize}
	
Biological structures as a rule have high complexity because the macrobehaviour is strongly influenced by the embedded microbehaviour. Mathematical modelling is a tool not only for describing biological processes and performing experiments \emph{in silico}. The behaviour of biomembrane is an excellent example how the microstructure (lipids) of a membrane has a direct impact on wave phenomena along the membrane. The analysis of the governing equation \eqref{EPT} (or Eq. \eqref{EPTdimensionless}) presented above demonstrates the richness of the model from the viewpoint of mathematical physics and opens the ways for physiological experiments concerning the properties of biomembranes. 

\section*{Acknowledgements}
This research was supported by the European Union through the European Regional Development Fund (Estonian Programme TK 124) and by the Estonian Research Council (projects IUT 33-24, PUT 434).


\newpage
\section*{References}
\begin{thebibliography}{41}
\expandafter\ifx\csname natexlab\endcsname\relax\def\natexlab#1{#1}\fi
\expandafter\ifx\csname url\endcsname\relax
  \def\url#1{\texttt{#1}}\fi
\expandafter\ifx\csname urlprefix\endcsname\relax\def\urlprefix{URL }\fi

\bibitem[{Ablowitz(2011)}]{Ablowitz2011}
Ablowitz, M.~J., 2011. {Nonlinear Dispersive Waves. Asymptotic Analysis and
  Solitons.} Cambridge Univ Press, Cambridge.

\bibitem[{Andersen et~al.(2009)Andersen, Jackson, and Heimburg}]{Andersen2009}
Andersen, S. S.~L., Jackson, A.~D., Heimburg, T., 2009. {Towards a
  thermodynamic theory of nerve pulse propagation.} Prog. Neurobiol. 88~(2),
  104--13.

\bibitem[{Appali et~al.(2012)Appali, {Van Rienen}, and Heimburg}]{Appali2012a}
Appali, R., {Van Rienen}, U., Heimburg, T., 2012. {A comparison of the
  Hodgkin-Huxley model and the soliton theory for the action potential in
  nerves}. In: Iglic, A. (Ed.), Adv. Planar Lipid Bilayers Liposomes, Vol. 16.
  Academic Press, Ch.~9, pp. 275--299.

\bibitem[{Berezovski et~al.(2013)Berezovski, Engelbrecht, Salupere, Tamm,
  Peets, and Berezovski}]{Berezovski2013a}
Berezovski, A., Engelbrecht, J., Salupere, A., Tamm, K., Peets, T., Berezovski,
  M., 2013. {Dispersive waves in microstructured solids}. Int. J. Solids
  Struct. 50~(11-12), 1981--1990.

\bibitem[{Boussinesq(1871)}]{Boussinesq1871}
Boussinesq, J., 1871. {Th{\'{e}}orie de l'intumescence liquide, applel{\'{e}}e
  onde solitaire ou de translation, se propageant dans un canal rectangulaire}.
  Comptes Rendus l'Academie des Sci. 72, 755--759.

\bibitem[{Christov et~al.(2007)Christov, Maugin, and Porubov}]{Christov2007}
Christov, C.~I., Maugin, G.~A., Porubov, A.~V., 2007. {On Boussinesq's paradigm
  in nonlinear wave propagation}. Comptes Rendus M{\'{e}}canique 335~(9-10),
  521--535.

\bibitem[{Dauxois and Peyrard(2006)}]{Dauxois2006}
Dauxois, T., Peyrard, M., 2006. {Physics of Solitons}. Cambridge University
  Press, Cambridge.

\bibitem[{Debanne et~al.(2011)Debanne, Campanac, Bialowas, Carlier, and
  Alcaraz}]{Debanne2011}
Debanne, D., Campanac, E., Bialowas, A., Carlier, E., Alcaraz, G., 2011. {Axon
  physiology}. Physiol. Rev. 91~(2), 555--602.

\bibitem[{Dodd et~al.(1982)Dodd, Eilbeck, Gibbon, and Morris}]{SaNWE1982}
Dodd, R., Eilbeck, J., Gibbon, J., Morris, H., 1982. {Solitons and nonlinear
  wave equations}. Academic Press Inc. LTD., London.

\bibitem[{Drazin and Johnson(1989)}]{Drazin1989}
Drazin, P., Johnson, R., 1989. {Solitons: and Introduction}. Cambridge
  University Press, Cambridge.

\bibitem[{Engelbrecht(1995)}]{engelbrecht1995}
Engelbrecht, J., 1995. {Beautiful dynamics}. Proc. Estonian Acad. Sci.
  Physics/Mathematics 1~(44), 108--119.

\bibitem[{Engelbrecht(1997)}]{Engelbrecht1997}
Engelbrecht, J., 1997. {Nonlinear Wave Dynamics. Complexity and Simplicity.}
  Kluwer, Dordrecht.

\bibitem[{Engelbrecht et~al.(2011)Engelbrecht, Salupere, and
  Tamm}]{Engelbrecht2011}
Engelbrecht, J., Salupere, A., Tamm, K., 2011. {Waves in microstructured solids
  and the Boussinesq paradigm}. Wave Motion 48~(8), 717--726.

\bibitem[{Engelbrecht et~al.(2015)Engelbrecht, Tamm, and
  Peets}]{Engelbrecht2015}
Engelbrecht, J., Tamm, K., Peets, T., 2015. {On mathematical modelling of
  solitary pulses in cylindrical biomembranes}. Biomech. Model. Mechanobiol.
  14, 159--167.

\bibitem[{Fornberg(1998)}]{Fornberg1998}
Fornberg, B., 1998. {A Practical Guide to Pseudospectral Methods}. Cambridge
  University Press, Cambridge.

\bibitem[{Fornberg and Sloan(1994)}]{ForenbergSloan94}
Fornberg, B., Sloan, D., 1994. {A review of pseudospectral methods for solving
  partial differential equations}. Acta Numer. 3, 203--267.

\bibitem[{Heimburg and Jackson(2007)}]{Heimburg2007}
Heimburg, T., Jackson, A., 2007. {On the action potential as a propagating
  density pulse and the role of anesthetics}. Biophys. Rev. Lett. 2, 57--78.

\bibitem[{Heimburg and Jackson(2005)}]{Heimburg2005}
Heimburg, T., Jackson, A.~D., 2005. {On soliton propagation in biomembranes and
  nerves.} Proc. Natl. Acad. Sci. U. S. A. 102~(28), 9790--5.

\bibitem[{Hindmarsh(1983)}]{ODE}
Hindmarsh, A., 1983. {ODEPACK, a Systematized Collection of ODE Solvers}.
  Vol.~1. North-Holland, Amsterdam.

\bibitem[{Hodgkin and Huxley(1952)}]{Hodgkin1952}
Hodgkin, A.~L., Huxley, A.~F., 1952. {A quantitative description of membrane
  current and its application to conduction and excitation in nerve}. J.
  Physiol. 117~(4), 500--544.

\bibitem[{Iwasa et~al.(1980)Iwasa, Tasaki, and Gibbons}]{Iwasa1980}
Iwasa, K., Tasaki, I., Gibbons, R., 1980. {Swelling of nerve fibers associated
  with action potentials}. Science 210~(4467), 338--339.

\bibitem[{Jones et~al.(2007)Jones, Oliphant, and Peterson}]{SciPy}
Jones, E., Oliphant, T., Peterson, P., 2007. {SciPy: open source scientific
  tools for Python}.

\bibitem[{Lautrup et~al.(2011)Lautrup, Appali, Jackson, and
  Heimburg}]{Lautrup2011}
Lautrup, B., Appali, R., Jackson, A.~D., Heimburg, T., 2011. {The stability of
  solitons in biomembranes and nerves.} Eur. Phys. J. E. Soft Matter 34~(6),
  1--9.

\bibitem[{Maugin(1995)}]{Maugin1995}
Maugin, G.~A., 1995. {On some generalizations of Boussinesq and KdV systems}.
  Proc. Estonian Acad. Sci. Phys. Math. 44, 40--55.

\bibitem[{Maugin(1999)}]{Maugin1999}
Maugin, G.~A., 1999. {Nonlinear Waves in Elastic Crystals}. Oxford University
  Press, Oxford.

\bibitem[{Maugin(2011)}]{Maugin2011}
Maugin, G.~A., 2011. {Solitons in elastic solids (1938-2010)}. Mech. Res.
  Commun. 38~(5), 341--349.

\bibitem[{Maurin(2015)}]{MaurinPHD}
Maurin, F., 2015. {Wave propagation in periodic buckled beams}. Ph.D. thesis,
  {\`{E}}cole Polytechnique F{\'{e}}d{\'{e}}rale de Lausanne.

\bibitem[{Maurin and Spadoni(2016{\natexlab{a}})}]{Maurin2016}
Maurin, F., Spadoni, A., 2016{\natexlab{a}}. {Wave propagation in periodic
  buckled beams. Part I: Analytical models and numerical simulations}. Wave
  Motion.

\bibitem[{Maurin and Spadoni(2016{\natexlab{b}})}]{Maurin2016a}
Maurin, F., Spadoni, A., may 2016{\natexlab{b}}. {Wave propagation in periodic
  buckled beams. Part II: Experiments}. Wave Motion.

\bibitem[{Mindlin(1964)}]{Mindlin1964}
Mindlin, R.~D., 1964. {Micro-structure in linear elasticity}. Arch. Ration.
  Mech. Anal. 16~(1), 51--78.

\bibitem[{Mueller and Tyler(2014)}]{Mueller2014}
Mueller, J.~K., Tyler, W.~J., 2014. {A quantitative overview of biophysical
  forces impinging on neural function.} Phys. Biol. 11~(5), 051001.

\bibitem[{Peets and Tamm(2015)}]{Peets2015}
Peets, T., Tamm, K., 2015. {On mechanical aspects of nerve pulse propagation
  and the Boussinesq paradigm}. Proc. Estonian Acad. Sci. 64~(3S), 331--337.

\bibitem[{Peets et~al.(2015)Peets, Tamm, and Engelbrecht}]{Peets2015a}
Peets, T., Tamm, K., Engelbrecht, J., 2015. {Numerical investigation of
  mechanical waves in biomembranes}. In: Elgeti, S., Simon, J.-W. (Eds.), Conf.
  Proc. YIC GACM 2015 3rd ECCOMAS Young Investig. Conf. 6th GACM Colloquium,
  July 20-23, 2015, Aachen, Ger. pp. 1--4.

\bibitem[{Peets et~al.(2016)Peets, Tamm, and Engelbrecht}]{Peets2016}
Peets, T., Tamm, K., Engelbrecht, J., 2016. {On the role of nonlinearities in
  the Boussinesq-type wave equations}. Wave Motion~(1), 1--7.

\bibitem[{Porubov(2003)}]{Porubov2003}
Porubov, A.~V., 2003. {Amplification of Nonlinear Strain Waves in Solids}.
  World Scientific, Singapore.

\bibitem[{Rayleigh(1876)}]{Rayleigh1876}
Rayleigh, L., 1876. {On waves}. Philos. Mag. 1~(1), 257--279.

\bibitem[{Salupere(2009)}]{Salupere2009}
Salupere, A., 2009. {The pseudospectral method and discrete spectral analysis}.
  In: Quak, E., Soomere, T. (Eds.), Appl. Wave Math. Springer Berlin
  Heidelberg, Berlin, pp. 301----334.

\bibitem[{Salupere et~al.(2002)Salupere, Peterson, and
  Engelbrecht}]{Salupere2002}
Salupere, A., Peterson, P., Engelbrecht, J., 2002. {Long-time behaviour of
  soliton ensembles . Part I –– Emergence of ensembles}. Chaos, Solitons
  {\&} Fractals 14, 1413--1424.

\bibitem[{Tamm and Peets(2015)}]{Tamm2015}
Tamm, K., Peets, T., 2015. {On solitary waves in case of amplitude-dependent
  nonlinearity}. Chaos, Solitons {\&} Fractals 73, 108--114.

\bibitem[{Tasaki(1988)}]{Tasaki1988}
Tasaki, I., 1988. {A macromolecular approach to excitation phenomena:
  mechanical and thermal changes in nerve during excitation}. Physiol. Chem.
  Phys. Med. NMR 20, 251--268.

\bibitem[{Zabusky and Kruskal(1965)}]{Zabusky1965}
Zabusky, N., Kruskal, M., 1965. {Interaction of ``solitons'' in a collisionless
  plasma and the recurrence of initial states}. Phys. Rev. Lett. 15~(6),
  240--243.

\end{thebibliography}
\end{document}